\documentclass[fleqn,usenatbib]{mnras}

\usepackage{newtxtext,newtxmath}
\usepackage[T1]{fontenc}

\DeclareRobustCommand{\VAN}[3]{#2}
\let\VANthebibliography\thebibliography
\def\thebibliography{\DeclareRobustCommand{\VAN}[3]{##3}\VANthebibliography}

\usepackage{graphicx}	% Including figure files
\usepackage{caption}
\usepackage{subcaption}
\newcommand{\be}{\begin{equation}}
\newcommand{\ee}{\end{equation}}

\newcommand{\ba}{\begin{eqnarray}}
\newcommand{\ea}{\end{eqnarray}}

\newcommand{\Alfven}{Alfv\'{e}n }

\newcommand\eg{{\it{e.g.\ }}}

\newcommand{\Bf}{{magnetic field}}
\newcommand{\Bfs}{{magnetic fields}}

\newcommand{\NS}{neutron star}

\newcommand{\mss}{magnetospheres}

\newcommand{\LC}{light cylinder}

\title{Radio afterglow of magnetars' giant flares}

\author[R. Mehta, M. Barkov and M. Lyutikov]{
Riddhi Mehta,$^{1}$\thanks{E-mail: mehta74@purdue.edu}
Maxim Barkov,$^{2}$\thanks{E-mail: barmv05@gmail.com}
and Maxim Lyutikov$^{1}$\thanks{E-mail: lyutikov@purdue.edu}
\\
$^{1}$Department of Physics and Astronomy, Purdue University, 525 Northwestern Avenue, West Lafayette,
IN 47907-2036, USA\\
$^{2}$Institute of Astronomy, Russian Academy of Sciences, Pyatnitskaya 48, 119017 Moscow, Russian Federation
}

\date{Accepted XXX. Received YYY; in original form ZZZ}

\pubyear{2021}

\begin{document}
\label{firstpage}
\pagerange{\pageref{firstpage}--\pageref{lastpage}}
\maketitle

\begin{abstract}
We develop a  model for the radio afterglow  of the giant flare of SGR 1806-20  arising due to the interaction of magnetically-dominated cloud, an analogue of  Solar Coronal Mass Ejections (CMEs), with the   interstellar medium (ISM). The  CME  is modeled as a spheromak-like configuration. The CME is first advected with the magnetar's  wind and later interacts with the ISM, creating a strong forward  shock and complicated backwards exhaust flow. Using three-dimensional magnetohydrodynamic simulations, we study various  relative configurations of the magnetic field  of the CME with respect to the ISM's \Bf. We show that the  dynamics of the forward shock mostly follows the  Sedov-Taylor blastwave, while the internal structure of the shocked medium  is considerably modified by the back flow, creating a multiple shock configuration.
We calculate  synthetic synchrotron emissivity maps and light curves using two assumptions: (i)   magnetic field  compression; (ii)   amplification of the  magnetic field 
at the shock. We find that models with magnetic field amplification account better for  the observed radio emission.  
\end{abstract}


%%%%%%%%%%%%%%%%%%%%%%%%%%%%%%%%%%%%%%%%%%%%%%%%%%

%%%%%%%%%%%%%%%%% BODY OF PAPER %%%%%%%%%%%%%%%%%%

\section{Introduction}
\label{sec:INTRO}

Soft gamma repeaters (SGRs) are a type of slowly spinning neutron stars exhibiting X-ray and $\gamma$-ray bursts. Together with Anomalous X-ray pulsars they constitute a class of magnetars  \citep{Thompson1995}, see reviews by  \cite{2008A&ARv..15..225M,2017ARA&A..55..261K}. Their emission is powered by the dissipation of \Bf, that may reach $B\approx10^{15}$ G.
Occasionally, SGRs produce  giant flares (GFs); GF from  SGR 1806-20  on  27 December 2004 is the most notorious \citep{Palmer2005,Hurley2005}. GFs 
consist of  bright, short initial spike peaking in soft $\gamma$-rays (with luminosities reaching  $10^{47}$ ergs/s and duration $\sim 100$ millisecond), followed by a longer and dimmer tail peaking in hard X-rays modulated by the SGR's rotational period and having luminosities upwards of $10^{44}$ ergs/s. \cite{Masada2010} suggest a model for magnetar giant flares based on solar flare/coronal mass ejection theory where the flare is the final outcome of prominence (loaded baryonic matter) eruption, triggered by explosive magnetic reconnection.

 %These GFs are believed to be powered by violent shearing and reconnection of the extreme magnetic fields on the SGR's surface \citep{Thompson1995,Thompson2001}. On 27 December 2004, SGR 1806-20 emitted one such GF \citep{Palmer2005,Hurley2005} with total (isotropic) flare energy about a hundred times higher than the two previously observed GFs. 

%[TWO FOLLOWING PARAGRAPHS DESCRIBING OBSERVATIONS NEED MORE EDITING]

A week after the GF of  SGR 1806-20, a bright radio afterglow was discovered \citep{Cameron2005,Gaensler2005}. The radio afterglow showed  one-sided, mildly collimated  and decelerating outflow with  initial expansion velocity of $\approx0.7c$
 \cite{Taylor2005,Granot2006}. The radio afterglow showed complicated behavior. At first, radio flux exhibited a moderate decay, $\sim t^{-1.5}$ to $\sim t^{-2}$ before 9 days after the giant flare,  after which it underwent an achromatic steepening, $\sim t^{-2.7}$ between $\sim9$ and $\sim25$ days \citep{Gaensler2005,Gelfand2005}. Starting at $\sim25$ days and peaking at $\sim33$ days, a rebrightening in the radio light curve was observed \citep{Gelfand2005}, followed by a slower decay, $\sim t^{-1.1}$.

The afterglow arises due to the interaction of an ejected blob (analogue of a Solar Coronal Mass Ejection, CME) with the surrounding medium.
Two contrasting suggestions of the composition of a CME  were proposed: matter versus \Bf-dominated ejections. In the former, the GF was accompanied by an ejection of  $>10^{24}$ grams of baryonic material (this requires slicing off several meters from the whole surface of a NS) with mildly relativistic velocities  \citep{Palmer2005,Gelfand2005,Taylor2005,Granot2006}.

In the latter model \citep[][``Solar flare paradigm'']{2006MNRAS.367.1594L,2015MNRAS.447.1407L}, GFs are magnetospheric events  ({\it not} crustal events),  driven by slow plastic evolution of the footpoints, qualitatively similar to Solar flares.  Magnetars bursts and flares are 
 driven by  unwinding  of the internal non-potential magnetic field via Hall (electron-MHD) drift \citep{RG,2013MNRAS.434.2480G,2014PhPl...21e2110W}. This 
 leads to 
 a slow build-up of magnetic energy  outside of the neutron star. For large magnetospheric currents,
 corresponding to 
 a large twist of the  external  magnetic field,
 magnetosphere  becomes dynamically unstable on the \Alfven crossing times scale
 of the inner magnetosphere. The ensuing reconnection processes, \eg\ mediated by  the tearing mode  in the strongly magnetized  plasma of magnetar \mss\  \citep{2003MNRAS.346..540L,2007MNRAS.374..415K,2019MNRAS.485..299R},  operates in a way qualitatively similar to the Sun. Since, in the ``Solar flare paradigm'', GFs are magnetospheric events, no large baryonic loading is expected.

In this paper  we  perform MHD simulations  of the interaction of the  ejected  light, nearly  magnetically-dominated  CME with the surrounding medium.
We outline the numerical set-up in section~\ref{sec:MHD}, describe results of the MHD simulations in section~\ref{sec:results}, and derive integrated synthetic synchrotron emissivity maps and light curves in section~\ref{sec:emiss_maps}. Discussion of results, conclusions, and future prospects are presented in section~\ref{sec:DIS}.%and~\ref{sec:CON}, respectively.

%present a model for the interaction of the outflow ejected during the SGR 1806-20 GF with its surroundings and derive synthetic synchrotron emissivity maps through three dimensional (3-D) magnetohydrodynamic (MHD) simulations in order to describe the observed radio emission. Our model (described in section~\ref{sec:AM}) considers that since magnetosphere explosion is expected to be light \citep{2006MNRAS.367.1594L}, material ejected during the GF is primarily in the form of magnetic fields, unlike previous suggestions of a heavy baryonic outflow. In particular, we model the outflow as a magnetically confined spheromak-like configuration described by \citet*{Gourgouliatos2010}. 

%%%%%%%%%%%%%%%%%%%%%%%%%%%%%%%%%%%%%%%%%%%%%

%\section{Modeling the radio emission associated with the giant flare}
\section{Magnetar's ejection as an expanding spheromak-like magnetic blob}
\label{sec:AM}

\subsection{Ejected magnetic blob in magnetar wind}
\label{frozen}

In this paper, we perform numerical calculations of radio afterglows of magnetar giant flares, modeled as relativistic magnetized explosions \citep{2006MNRAS.367.1594L}. According to the ``The Solar model of magnetars'',  the explosions are magnetospheric-driven \citep[not crustal driven ][]{Thompson1995} events
\citep[see also][]{2012MNRAS.427.1574L,2015MNRAS.447.1407L}. In the case of magnetospheric release of energy, one does not expect substantial baryonic loading of the expelled fireball.

 Initially, at the moment of launching, the blob is over-pressurized (both due to pair plasma and the internal  \Bf) and has linear momentum implanted during the ejection.   As the magnetic cloud expands, its pair density falls by many orders of magnitude forming light, magnetically dominated blob. 
Initially,  the  magnetic cloud is topologically connected to the star, but  reconnection 
at the footprints disconnects it.
We expect the blob material to be slightly different (a bit higher density) from the wind due to pair freeze-out.

The internal structure of the blob frozen into the wind may be approximated as a spheromak-like configuration, a spherical  nearly force-free configuration of linked poloidal and toroidal \Bfs.  In the case of the Sun both spheromaks and flux ropes are  used to model expanding structures  \citep{2000JGR...105.2375L,2000ApJ...545..524C,1998JGR...10323717V} \citep[though flux ropes are generally preferred, ][]{1995JGR...10012293F}.  In the case of the Sun the expansion from the surface to the Earth radius is  a factor of $\sim 100$.  In our case the expansion form the surface of a NS to the region of interaction with the ISM is more than 10 orders of magnitude. As a result the ejected material is likely to be disconnected. Hence we accept the spheromak as a model of the ejections.

The  pre-explosion wind is expected to accelerate linearly from the \LC,\  $\Gamma_ w \propto r/R_{LC}$ ($R_{LC}$ is the radius of the \LC). The over-pressurized blob is also expected to accelerate with the same scaling \cite[similar to fireball model of GRBs][]{1986ApJ...308L..43P}. Thus, radial expansion of the blob and the wind are very similar - they virtually do not interact. 
The implanted linear momentum changes this picture only slightly: due to relativistic freeze-out  in radially accelerated flow, the relative velocity of the blob with respect to the wind quickly becomes non-relativistic. Pressure balance is then  established: the blob is frozen into the wind, and is advected with the wind.

As the wind expands and its rest-frame  \Bf\ decreases, the trapped blob will adjust to the changing confining pressure:  the spheromak will expand
(in fact, in accelerating flow the spheromak may even become causally disconnected internally).
 \cite{GV2010,2011SoPh..270..537L} discuss the structure of expanding force-free structures. Expanding spheromak develops internal velocity and currents to compensate for different scalings of toroidal and poloidal components of the magnetic field.

\begin{figure*}	
\centering
%\begin{subfigure}[b]{0.45\textwidth}
   \includegraphics[width=18cm]{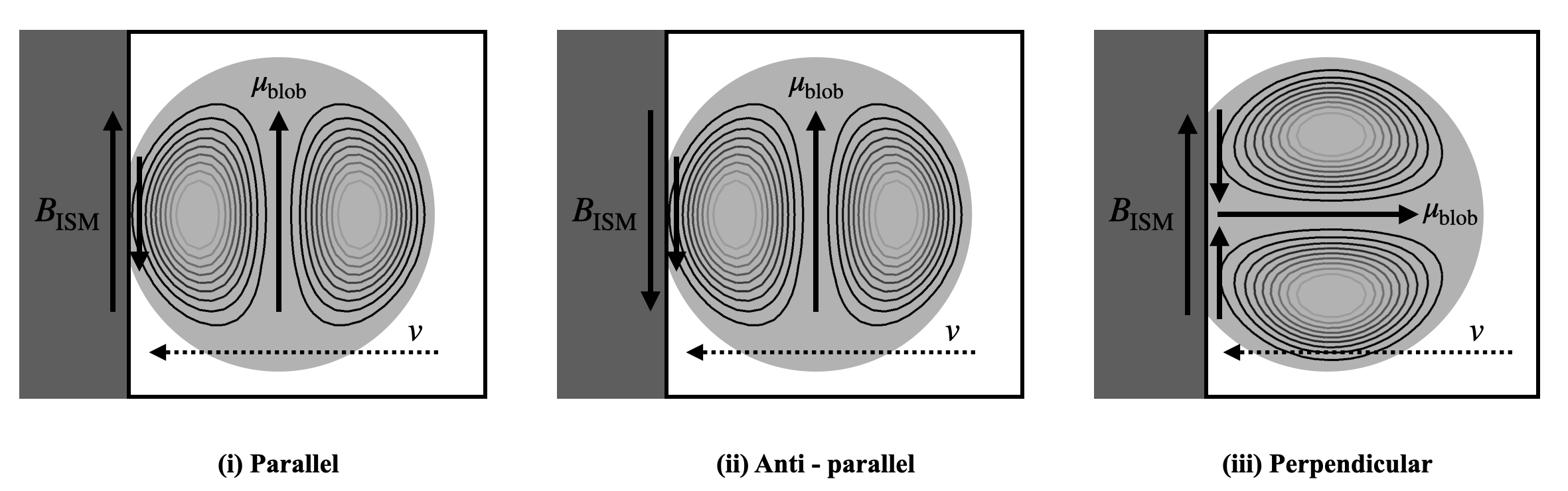}
 %  \caption{}
%   \label{fig:setup}
%\end{subfigure}
%\begin{subfigure}[b]{0.45\textwidth}
   \includegraphics[width=18cm]{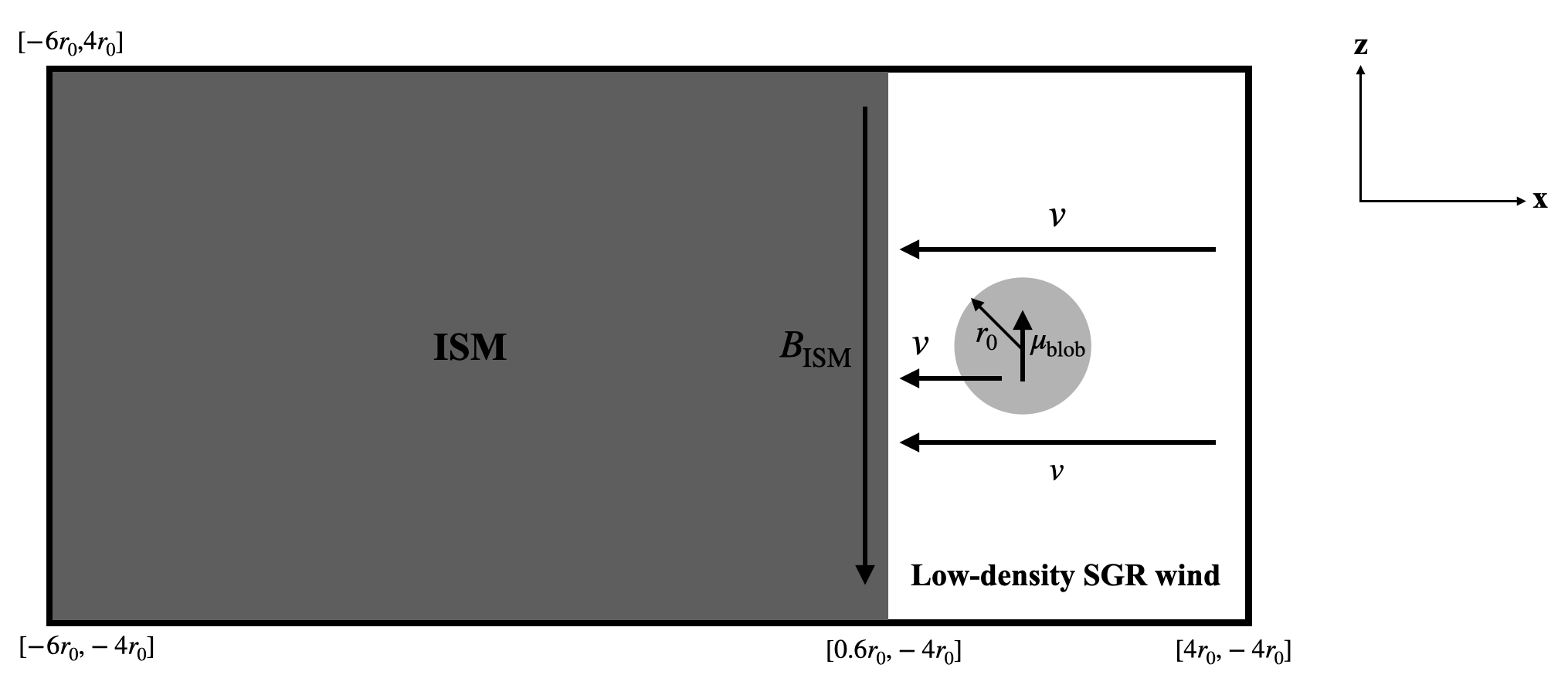}
  %  \caption{}
%    \label{fig:setup}
%\end{subfigure}
 \caption{Top panel: Schematic representation of the set-ups used to analyze  interaction of the blob with the ISM with three different orientations. The blob is enlarged to show the concentric flux surfaces that define the magnetic field structure within it. Bottom panel: Schematic representation of the anti-parallel set-up used to assess shock properties after the blob-ISM interaction with minimum effects of magnetic reconnection. $v$ is the velocity of the blob and SGR wind moving towards the ISM (see section~\ref{sec:num_set} for details).}
\label{fig:setup}
\end{figure*}

Thus, we expect that by the time the wind starts interacting with the external medium \citep[\eg at the bow-shock][]{2019MNRAS.484.4760B}, the ejected blob would have expanded to the size of the  fraction of the radius.  Before interacting with the ISM, the wind passes through the reverse shock. We assume that the blob identity is not destroyed (we {\it do} see an ejection!). A slightly denser blob will then start interacting with the ISM. Our simulations begin here, figure~\ref{fig:setup}.

\subsection{Magnetic blob -  magnetized ISM interaction}

We model the interaction of a magnetic blob with the ISM, figure~\ref{fig:setup}. Magnetic blob is modeled as a  spheromak-like configuration. Spheromaks are simply connected magnetically confined spherical structures which are spontaneously created due to plasma relaxation \citep{2018mhss.book.....B}. 
A confined classical spheromak requires surface  current.  In order to avoid complication with internal  resistive effects as the spheromak enters the ISM,
we use a slightly modified magnetic field structure of \citet{Gourgouliatos2010}  with vanishing toroidal and poloidal \Bfs\ on the boundary; such a configuration is slightly non-force-free (see appendix \ref{sec:GS}).

%We model the radio emission using the magnetic field structure of \citet{Gourgouliatos2010} namely a structure demanding vanishing magnetic field on the surface due to unmagnetized external plasma.

%Hence, such configurations can be thought to form in magnetospheres of magnetars where there is ample plasma and magnetic field. In the following section, we describe the analytical model used to model the radio emission associated with the giant flare from magnetar SGR 1806-20.

Interaction of the wind of the fast moving \NS\ with the ISM is a complicated 3D MHD problem \citep{2019MNRAS.484.4760B}. 
 Interaction of the expelled spheromak with the ISM adds further complications. For examples, details will depend on the relative  directions of the pulsar's velocity and ejections vector. As a simplified model we consider planar case, neglecting the curvature wind-ISM boundary  and the  possible obliqueness of the ejection. Thus, we assume that the size of the spheromak is somewhat smaller than radius of curvature of the ISM-wind boundary \citep[expected be $\sim 10^{16}$ cm, \eg][]{2006ARA&A..44...17G}, and that the interaction is near the apex of the bow-shock (so that  the impact is normal.)

In our simulations a spheromak-like 
magnetic blob (light grey in figure \ref{fig:setup}),  embedded within a low-density SGR wind, impacts on the ISM (dark grey). Internal \Bf\ within the blob may have three generic orientations  with respect to the   orientation of ISM's uniform magnetic field $B_{\text{ISM}}$.  The blob's magnetic moment $\mu_{\text{blob}}$  can be (i) parallel (ii) anti-parallel (iii) perpendicular to the external field (panel \ref{fig:setup}). These three configurations are expected to produce somewhat different magnetic interactions between the ISM's and the blob's \Bfs\ (see section \ref{sec:recon_results}).
 
\section{Three-dimensional MHD Simulations}
\label{sec:MHD}

\subsection{Numerical set-up}
\label{sec:num_set}

We perform 3-D MHD simulations to study the interaction of the aforementioned magnetic blob moving along with the low-density SGR wind, with an external ISM and consequently describe the radio nebula associated with the 2004 GF. Our aim is to simulate a strong shock (Mach number $\sim 10-20$) that would result from the interaction of the fast moving blob and wind (in the ISM's reference frame) with the ISM with low sound speed. A very strong shock is expected when material and magnetic fields ejected from the SGR's flare hit the ISM with velocity $\sim0.3c$ as observations suggest. It is the synchrotron emission from this shock that we hypothesize to have been observed as the radio emission and whose properties we wish to analyze for comparison with the observed spectrum.

The simulations are performed using a 3-D Cartesian geometry using the \textit{PLUTO} code\footnote{http://plutocode.ph.unito.it/index.html} \citep{Mignone2007}. \textit{PLUTO} is a modular Godunov-type code entirely written in C, intended mainly for astrophysical applications and high Mach number flows in multiple spatial dimensions and designed to integrate a general system of conservation laws
\begin{equation}
    \frac{\partial \mathbfit{U}}{\partial t} = -\nabla \cdot \mathbfit{T}(\mathbfit{U}) + \mathbfit{S}(\mathbfit{U})
    \label{tensor}
\end{equation}
$\mathbfit{U}$ is the vector of conservative variables and $\mathbfit{T}(\mathbfit{U})$ is the matrix of fluxes associated with those variables. Our ideal MHD set-up does not use any source terms and $\mathbfit{U}$ and $\mathbfit{T}$ are
\begin{equation}
\mathbfit{U} = \begin{pmatrix} \rho\\ \mathbfit{m}\\ \mathbfit{B}\\ E  \end{pmatrix},
\mathbfit{T}(\mathbfit{U}) = \begin{bmatrix} \rho \mathbfit{v} \\ \mathbfit{m v} - \mathbfit{BB} + p_t \mathbfit{I} \\ \mathbfit{vB - Bv} \\ (E + p_t)\mathbfit{v} - \mathbfit{(v B)B} \end{bmatrix}^\text{T}
\label{U&T}
\end{equation}
$\rho, \mathbfit{v}$ and $p$ are density, velocity and thermal pressure. $\mathbfit{m} = \rho \mathbfit{v}$, $\mathbfit{B}$ is the magnetic field and $p_t = p + |{\mathbfit{B}}|^{2}/2$ is the total (thermal + magnetic) pressure, respectively. Magnetic field evolution is complemented by the additional constraint $\nabla \cdot \mathbfit{B}=0$. Total energy density $E$
\begin{equation}
    E = \frac{p}{\Gamma - 1} + \frac{1}{2}\left(\frac{|\mathbfit{m}|^{2}}{\rho} + |{\mathbfit{B}}|^{2}\right)
    \label{energy_density}
\end{equation}
along with an isothermal equation of state $p = c_s^2 \rho$ provides the closure. $\Gamma$ and $c_s$ are the polytropic index and isothermal sound speed, respectively. The plasma is approximated as an ideal, non-relativistic adiabatic gas, one particle species with polytropic index of 5/3. \textit{PARABOLIC} interpolation, a third-order Runge-Kutta approximation in time, and a Harten-Lax-Van Leer approximate Riemann solver \citep{Miyoshi2005} are used to solve the above ideal MHD equations. Outflow boundary conditions are applied in all three directions. 

%\subsection{Magnetic reconnection}
%\label{sec:mag_recon}

%\subsection{Physical set-up}
%\label{sec:recon_setup}

We performed a short, low-resolution simulation to resolve the magnetic field structure at early times and assess the effects of current sheet formation and magnetic reconnection when the magnetic blob and low-density SGR wind interact with the ISM with different magnetic field orientations with respect to the blob's magnetic moment $\mu_{\text{blob}}$. Panel~\ref{fig:setup} shows a schematic of the physical set-up with parallel, anti-parallel and perpendicular orientations of $B_{\text{ISM}}$ with respect to $\mu_{\text{blob}}$. Specifically, we run the parallel and anti-parallel cases to analyze effects of reconnection and expect that the anti-parallel case, where $B_{\text{ISM}}$ and magnetic field at the blob's left edge are aligned, would be favorable to minimize reconnection effects and analyze the shock properties to explain the radio emission from the GF.

The size of the domain is $x \in [-2.4r_0, 4r_0]$, $y \in [-2.4r_0, 2.4r_0]$ and $z \in [-2.4r_0, 2.4r_0]$ where $r_0$ is the radius of the blob. Uniform resolution is used throughout the computational domain with total number of cells $N_{\rm X}=N_{\rm Y}=N_{\rm Z}$ = 256. The blob's initial magnetic field is defined by~\ref{eq:Br},~\ref{eq:Btheta} and~\ref{eq:Bphi}. We prefer low plasma-$\beta$ (high magnetization) of the blob so that effect of magnetic field is captured. These requirements make the following choice of initial parameters justified: an initial velocity of the blob and SGR wind in the negative $x$ direction \mathbfit{v} = -100$\hat{x}$, ISM pressure $p_\text{ISM} = 0.5$ and ISM density $\rho_\text{ISM} = 0.25$ giving the ISM sound speed $c_s$= 1.82, uniform ISM magnetic field \mathbfit{$B_\text{ISM}$} = -0.25$\hat{z}$, giving $\beta_{ISM}=16$. The wind pressure is  $p_\text{wind} = 0.5$, wind density $\rho_\text{wind} = 3\times10^{-5}$, $r_0 = 50, \beta = 2$  (not very small value of beta of the blob used in the simulations comes from purely numerical limitations) and blob density $\rho_\text{blob}=0.1\rho_\text{ISM}$. All quantities are given in code units which are normalized $cgs$ values
\begin{equation}
    \rho = \frac{\rho_{cgs}}{\rho_n},
    v = \frac{v_{cgs}}{v_n},
    p = \frac{p_{cgs}}{\rho_n v_n^2},
    B = \frac{B_{cgs}}{\sqrt{4 \pi \rho_n v_n^2}}
     \label{code_units}
\end{equation}
$\rho$, $v$, $p$ and $B$ are density, velocity, pressure and magnetic field. Time is given in units of $t_n=L_n/v_n$. The normalization values used are $\rho_n = 1.67 \times 10^{-24} \text{gr/cm}^3$, $L_n = 1.5 \times 10^{13}$ cm and $v_n = 10^5$ cm/s. We describe results of this analysis later in section~\ref{sec:recon_results}.

To model the radio emission from the GF, we choose the anti-parallel orientation ((ii) of panel~\ref{fig:setup}), with the ISM magnetic field aligned with the magnetic field at the blob's left edge and capture the shock dynamics arising from the interaction of magnetized blob and low-density SGR wind with external ISM in high resolution for longer times. Cases (i) and (ii) of panel~\ref{fig:setup} evolve almost identically which will be shown in section~\ref{sec:recon_results}. A 2-D ($xz$ plane) schematic of the anti-parallel set-up is shown in panel~\ref{fig:setup}. The size of the domain is $x \in [-6r_0, 4r_0]$, $y \in [-4r_0, 4r_0]$ and $z \in [-4r_0, 4r_0]$ where $r_0$ is the radius of the blob. Uniform resolution is used throughout the computational domain with total number of cells $N_{\rm X}=N_{\rm Y}=N_{\rm Z}$ = 780. The ISM extends from $-6r_0$ to $0.6r_0$ and the low-density cavity extends from $0.6r_0$ to $4r_0$ along the $x$ direction. We center the blob at $[2r_0,0,0]$ oriented such that $\mu_{\text{blob}}$ and $B_{\text{ISM}}$ are anti-parallel. The blob (light grey) is embedded within the SGR's low-density wind and both move towards the stationary ISM (dark grey) with velocity $v$. The initial parameters of the blob, ISM and SGR wind and the normalizing values are the same as described above. 

\subsection{Theoretical expectations}
\label{sec:ZR_DIS}

The key physical inputs are: a blob (projectile) creates a bulk moving and  over-pressurized region  (implanted momentum and energy). Both, the implanted momentum and the implanted energy ``push'' the forward shock (FS). But in contrast to the spherical Sedov explosion (where opposite parts of the flow ``push'' against each other, thus not conserving the absolute value of momentum), in the case of projectile hitting the ground, energy and momentum can be lost to the backward flow. This energy/momentum loss is complicated: parts of the post-shock flow close to the FS move with nearly the velocity of the FS ($3/4$ of the shock velocity), and thus with supersonic velocities in the frame of the shell. Further downstream, bulk velocities decrease, become subsonic, and start to ``feel'' the absence of the back confinement  - the resulting high pressure accelerates the flow backwards, toward the wind, forming complicated exhaust flows.  Because of high initial pressure, the exhaust flows  become supersonic and form a series of shocks as seen in figure \ref{fig:vx3_20}.

The implanted momentum and the pressure loss through backward exhaust flow act in the opposite way:  implanted momentum increases the FS velocity, while the pressure loss from the bulk drains the energy, and hence leads to the slowing down of the shock.

For the dynamics of the FS, 
the problem under consideration is somewhat similar to the classical  problem of projectile hitting the  ground \citep{ZelRai1967,Whitham}: energy and momentum are implanted.  
Also, recall that in the bulk plasma the magnetization is low  (beta-parameter in the ISM is $\beta_{ISM} = 25$, so that \Bfs\ in the ISM do not affect the flow considerably. But they are of primary importance for the production of radiation. As \cite{ZelRai1967} discuss, the  resulting shock dynamics is limited between the two cases of energy and  momentum conservation.  For pure energy injection with $E_0$, we expect that the scaling of the shock's radius follows the Sedov solution \cite{Sedov}
\be
R_E \sim \left( \frac{E_0 t^2} {\rho} \right)^{1/5}
\ee
For pure momentum injection with $P_0 \sim \sqrt{M_0 E_0}$ ($M_0$ is the  mass of the blob/projectile),  we expect that the scaling of the shock's radius follows the Kompaneets solution \cite{1960SPhD....5...46K}
\be
R_P \sim \left( \frac{P_0 t} {\rho} \right)^{1/4}
\ee
(We stress that the resulting dynamics is not self-similar, but  is ``bracketed''  between these two self-similar solutions). Closeness to any of the above solution depends on the non-dimensionless parameter $E_0/P_0$, and hence cannot be easily quantified.\\
Defining Sedov radius and time as
\be
R_S = \left( \frac{M_0}{\rho}  \right)^{1/3}
\ee
\be
t_S = \frac{ M_0^{5/6}}{E_0^{1/2} \rho^{1/3}}
\ee
the momentum injection dominates the FS dynamics at times and radii $\leq t_S, \, R_S$,
\be
\frac{R_E}{R_P} = \left (  \frac{ t}{t_S} \right)^{3/20} 
\ee

%Defining Sedov radius  and time as
%\ba &&
%R_S = \left( \frac{M_0}{\rho}  \right)^{1/3}
%\nn &&
%t_S = \frac{ M_0^{5/6}}{E_0^{1/2} \rho^{1/3}}
%\ea
%the momentum injection dominates the FS dynamics at times and radii $\leq t_S, \, R_S$,
%\be
%\frac{ R_E}{R_P} = \left (  \frac{ t}{t_S} \right)^{3/20} 
%\ee

Though the overall dynamics is not self-similar, 
we may use the self-similar scalings  $p\sim M^{-n}\sim R^{-3n}$ where $p$ is the shock pressure, $M$ is the mass encompassed by the shock wave and $R$ is the shock radius, with an exponent $n$ that may vary in time.
Using   the time dependences of shock pressure and radius (deduced in section~\ref{sec:p_v_R} and figure~\ref{fig:p_R_TE}),  we find $n\approx1.05$. This exponent value of shock attenuation is in excellent agreement to the proposed range $1 < n < 1.275$ for a self-similar concentrated impact for $\gamma=5/3$. The lower limit corresponds to an energy conserving shock and the upper limit corresponds to momentum conservation. Thus, the FS's self-similarity is intermediate between energy conserving and momentum conserving regimes. It is important to further note that the concentrated impact is closer to a point explosion in an infinite medium.
%%%%%%%%%%%%%%%%%%%%%%%%%%%%%%%%%%%%%%%%%%%%%

\section{Results}
\label{sec:results}

\begin{figure*}	
\includegraphics[width=18cm]{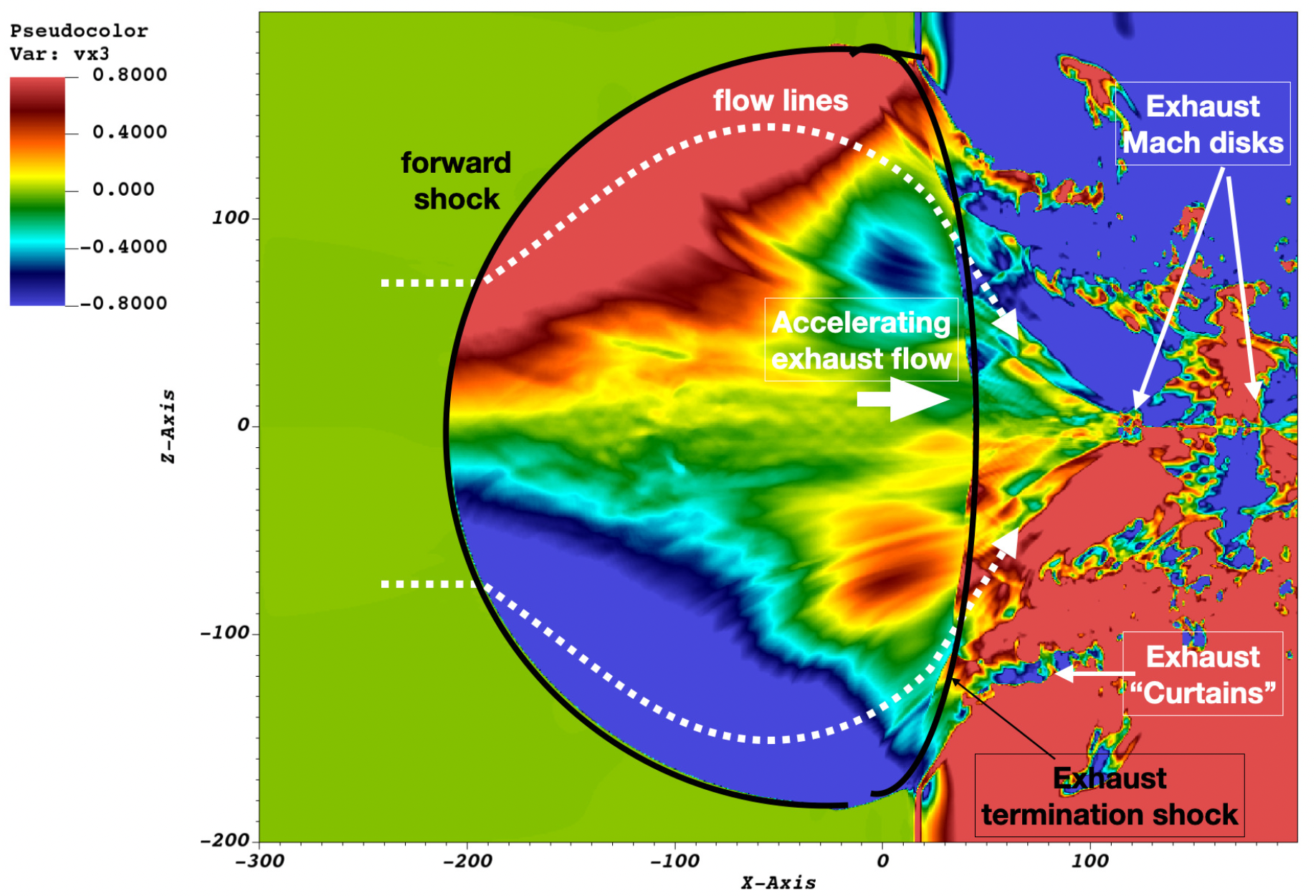}
\caption{ Key features of the flow annotated. Plotted is the $z$  (vertical) component  of velocity at  fairly advanced time of $t=20$. Upstream plasma (green color) is stationary. The forward shock (thick black line on the left)  is, generally, oblique, inducing  post-shock $z$ flow up/down in the upper/lower parts. In the bulk of the shocked material, soon after the FS, the rarefaction wave induces  ``exhaust flow'' towards the back end (large white arrow). The exhaust flow becomes supersonic and terminates at the ``exhaust
termination shock'' (thick black line near $x=0$). The post-exhaust
termination shock flow, collimated towards the symmetry axis,  experiences further  shocks at ``exhaust
Mach disks''.  Near the regions where the FS intersects with the boundary, high post-shock pressure launches ``exhaust curtains'' back in the wind.}
\label{fig:vx3_20}
\end{figure*}

We performed a number of  simulations with different resolutions - low ($256^3$) and high ($780^3$). In this section, we discuss various, quite complicated aspects of the  resulting flow. The salient features  are highlighted in figure \ref{fig:vx3_20}, where we plot the $z$   (vertical) component of the velocity at late times,  $t=20$ in code units. 

The flow is first shocked at the forward shock. This creates a region of high pressure. For spherical explosion, this region  of  high pressure has only one way to expand - by driving the FS. In the impact case, there is a second exit: forming a back flow. As a result, the high post-FS pressure is released both, through driving the FS, and through generation of the backward flow.

The overall dynamics of the FS is fairly simple: in the strong shock limit, it approximately  follows a self-similar solution, though with not well-defined parameters, bounded by the limits of energy and momentum conservation (section \ref{sec:ZR_DIS}).
 The back flow is more complicated. As the high post-FS pressure is converted into backward  motion, the flow becomes supersonic. As a result, a  termination shock forms (back black curve in figure \ref{fig:vx3_20}).  At this termination shock the flow is deflected toward the symmetry axis; it overshoots the pressure balance and forms repetitive Mach disks, similar to when a jet air plane flows at low altitudes (when the post-nozzle pressure does not match the ambient pressure). 
 
Additional features appear at the intersection of the FS and  the boundary between ISM and magnetar wind. High post-shock pressure drives an expansion curtain back into the wind, in a way similar to charged explosions hitting the ground, and asteroid impacts \citep{1989icgp.book.....M}.
 
In addition to the hydrodynamically complicated flow, discussed in detail in sections \ref{sec:p_v_R} and \ref{sec:SS}, the presence of \Bf\ in the blob, and its interaction with the external \Bf\ adds few complications to the classical problem of charged projectile impact (section \ref{sec:ZR_DIS}).

In the following discussion, we describe the high-resolution numerical results of the model described in section \ref{sec:num_set} and panel \ref{fig:setup}, namely 2-D slices of pressure, magnetic field, and $x$ component of velocity at $t=5, 10, 15$ and 20 as well as time evolution of pressure and radius of the FS.

%%%%%%%%%%%%%%%%%%%%%%%%%%%%%%%%%%%%%%%%%%%%%

\subsection{Pressure, velocity, and radius of the shock}
\label{sec:p_v_R}

\begin{figure*}	
   \includegraphics[width=18cm]{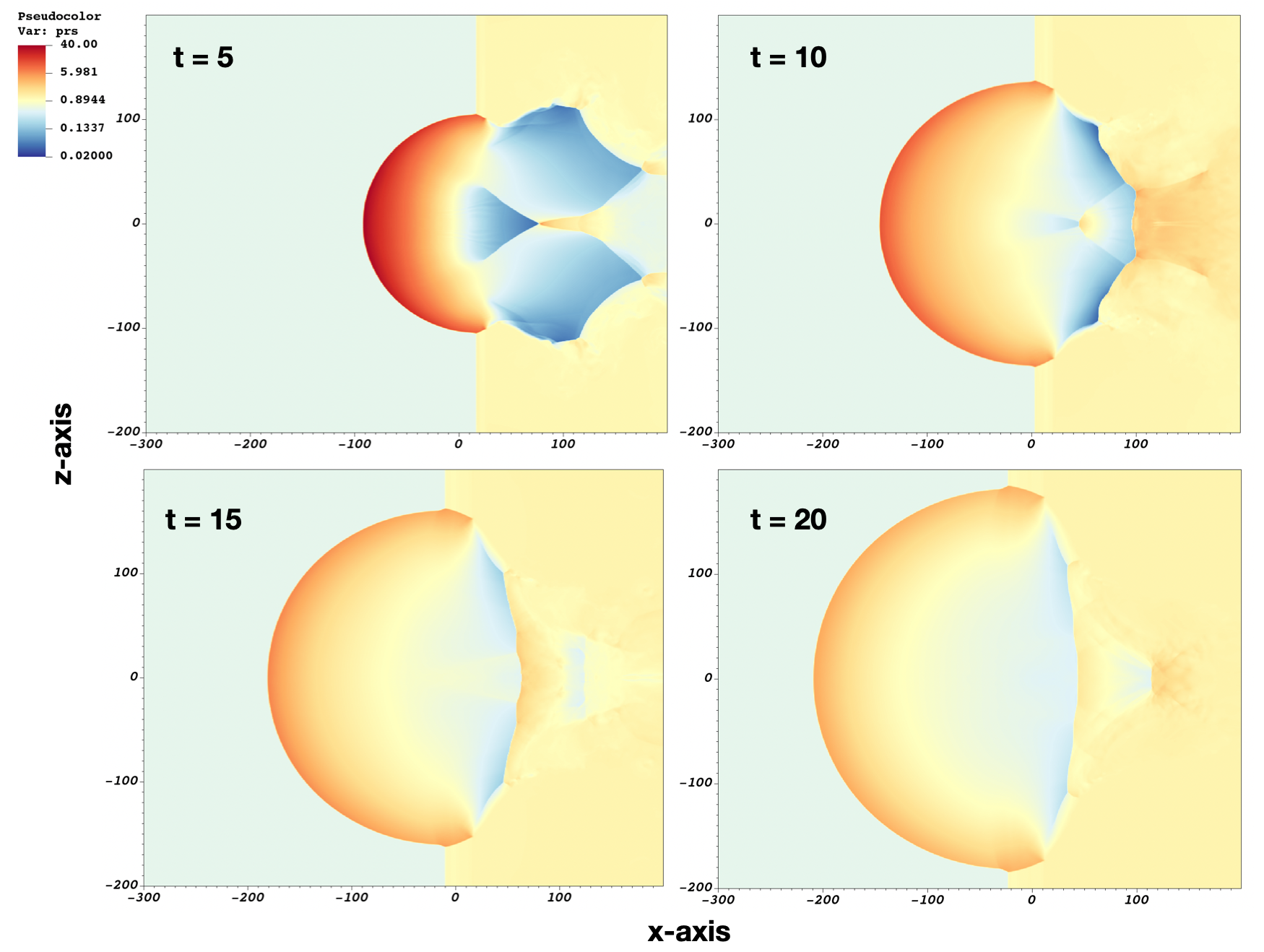}
    \caption{Slices in the $xz$ plane of the high-resolution MHD simulation of the interaction of radio blob with external ISM captured at $t=5,10,15$ and 20 in code units. Colors indicate pressure in log scale.}
    \label{fig:prs_map}
\end{figure*}

\begin{figure*}	
   \includegraphics[width=18cm]{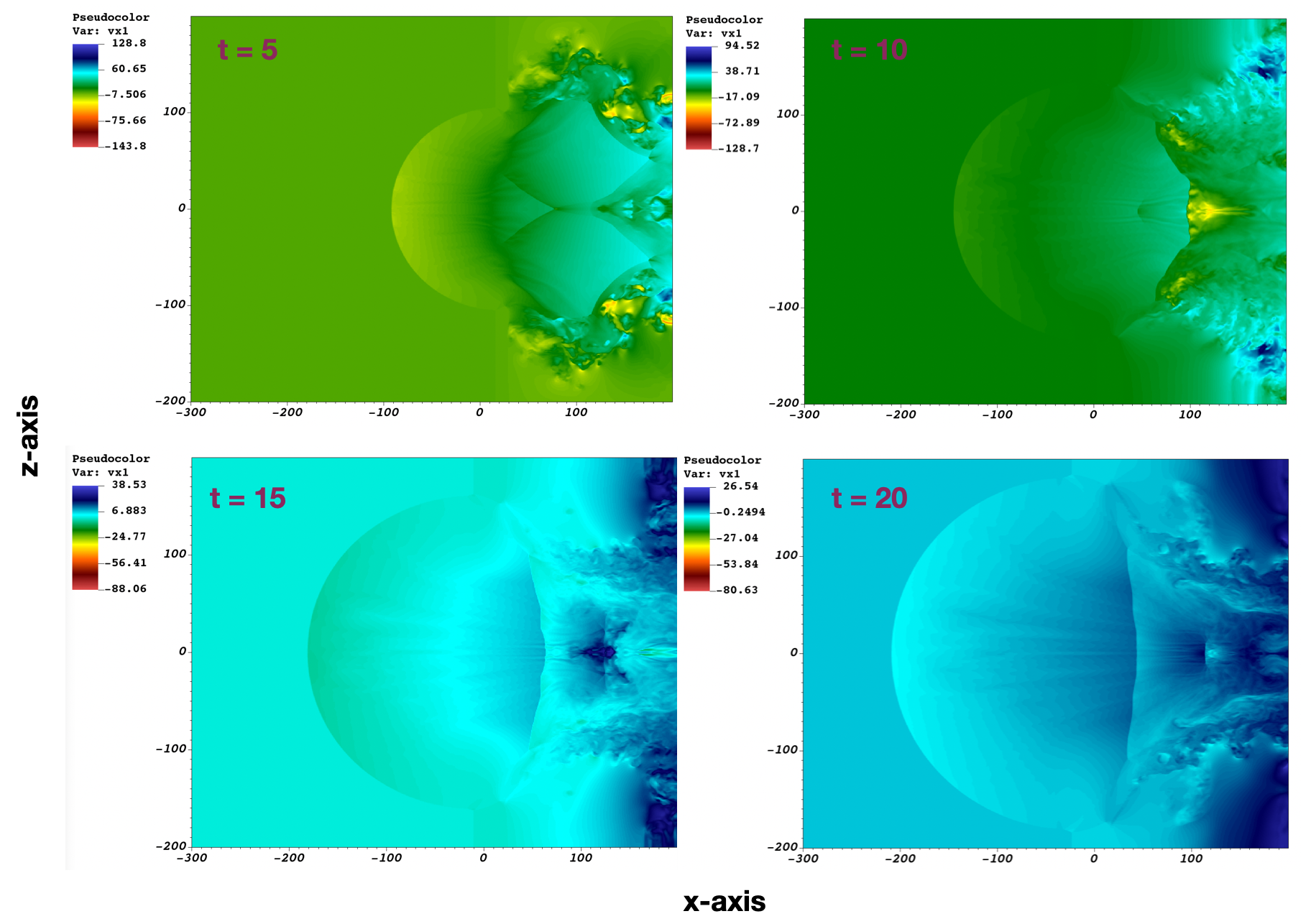}
    \caption{Same as figure~\ref{fig:prs_map} but colors indicate $x$ component of velocity.}
    \label{fig:vx_map}
\end{figure*}

At $t=0$, the blob and low-density wind start moving to the left and hit the ISM boundary. This interface between the ISM and cavity wind is where the shock originates and the FS continues to propagate towards the left. We capture the FS until $t=20$ after which it begins to exit the domain. Figures~\ref{fig:prs_map} and~\ref{fig:vx_map} %{\color{red} we need to expand the description that we see on the figures, describe them individually. Do we need so many plots? why not plot several lines on one plot? I think plots with t=15 do not bring a new information. } 
are pressure and $x$ component of velocity $vx1$ projected in 2-D ($xz$ plane) at $t=5,10,15$ and 20 where significant changes to the shock's morphology can be observed.

At $t=5$, after initial numerical fluctuations settle, a spherical FS is seen traveling to the left. At the same time, two wing-shaped low-pressure regions develop to the right of the FS into the cavity penetrated by a weak pressure peak just beginning to form. This is the recollimation shock being ejected out of the blob. The pressure depressions can be understood by noticing that at $t=5$, material behind the FS is being blown to the right with much greater speeds than that of the material moving to the left.

At $t=10$, as the FS penetrates deeper into the ISM, the recollimation shock is followed by a third pressure bump developing at the edge of the domain. This last shock is a result of external pressure in the wind zone as it thrashes against the recollimation shock and ISM. By this time, the FS is weakened by almost a third due to the strong ``exhaust flows'' described previously. Direction of moving material within the domain can be seen from its velocity map - FS moving to the left is followed by material moving right. This is followed by a weak recollimation shock and a powerful wind shock moving left at speed greater than the rest of the material.

Final stages of the shock at $t=15$ and $t=20$ are dominated by ``exhaust'' moving to the right at the edge of the domain caused due to reflected pressure waves. The recollimation and wind shocks are stronger but the FS is weakened nine-fold. The concave boundary prominent just to the right of the ISM-cavity interface is formed because material propagating to the right just behind the FS intercepts the wind propagating to the left at greater pressure than the wind's pressure. This exhaust carries away pressure, hence energy from the bulk, thus slowing down the FS (by $\sim 1/2$ from $t=5$ to $t=20$) as evident from the $vx1$ map.

\begin{figure}
\includegraphics[width=\columnwidth]{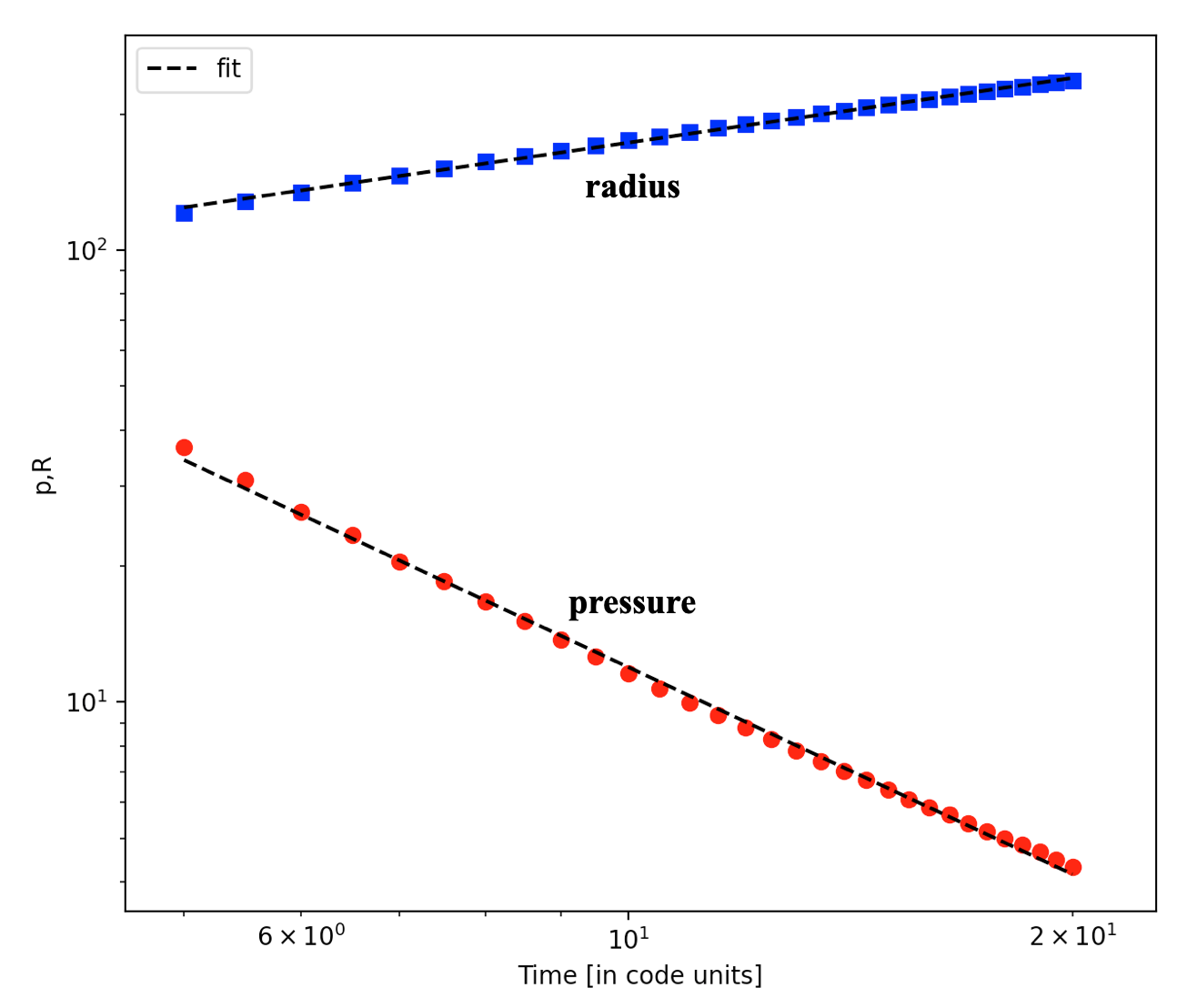}
    \caption{Time evolution of shock pressure $p$ (red circles) and radius $R$ (blue squares) in log-log scale. A linear fit through the data (black dashed lines) gives $p(t) \propto t^{-1.52}$ and $R(t) \propto t^{0.48}$.}
    \label{fig:p_R_TE}
\end{figure}

Time evolution of pressure and radius of the FS are depicted in figure~\ref{fig:p_R_TE}. We plot the peak pressure along $z=0$ and radius of the shock, that is, distance traveled by the point where pressure peaks along $z=0$, from $t=5$ to $t=20$ in steps of 0.5 in log-log scale. A linear fit to the plots gives power laws for both, pressure and radius of the shock: $p(t) \propto t^{-1.52}$ and $R(t) \propto t^{0.48}$. 

%\begin{figure*}
%   \includegraphics[width=18cm]{prs_radial.png}
%    \caption{Pressure profiles along $z=0$ plotted at the same four time instants as figure~\ref{fig:prs_map}. At $t=20$, forward shock to the left, recollimation shock in the middle and ``wind'' shock to the right are clearly visible. The final stages of shock are dominated by ``exhaust flows'', weakening the shock by almost $1/9$.}
%    \label{fig:prs_radial}
%\end{figure*}

%\begin{figure*}	
%   \includegraphics[width=18cm]{vx1_radial.png}
%    \caption{$vx1$ profiles plotted along $z=0$ at the same four time instants as figure~\ref{fig:prs_map}. A significant back flow of material is seen at $t=5$. At $t=10$ the ``wind'' shock moving left pressurizes the recollimation shock. The final stages of shock are dominated by ``exhaust flows'' exiting the domain.}
%    \label{fig:vx1_radial}
%\end{figure*}

%%%%%%%%%%%%%%%%%%%%%%%%%%%%%%%%%%%%%%%%%%%
\subsection{Properties of the  forward shock}
\label{sec:SS}

\begin{figure*}	
   \includegraphics[width=18cm]{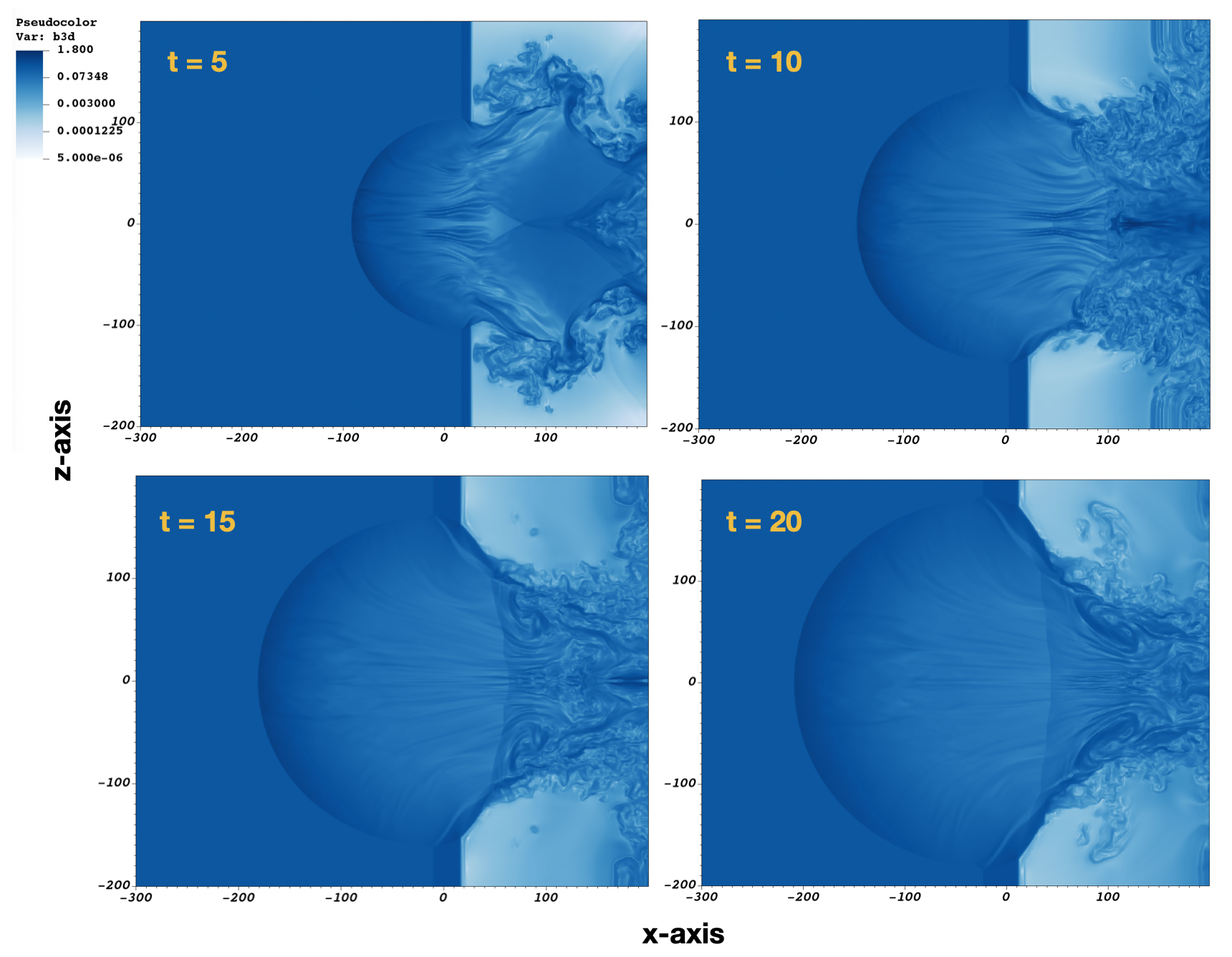}
    \caption{Same as figure~\ref{fig:prs_map} but colors indicate net magnetic field in log scale.}
    \label{fig:b3d_map}
\end{figure*}

\begin{figure*}	
\centering
%\begin{subfigure}[b]{0.475\textwidth}
   \includegraphics[width=18cm]{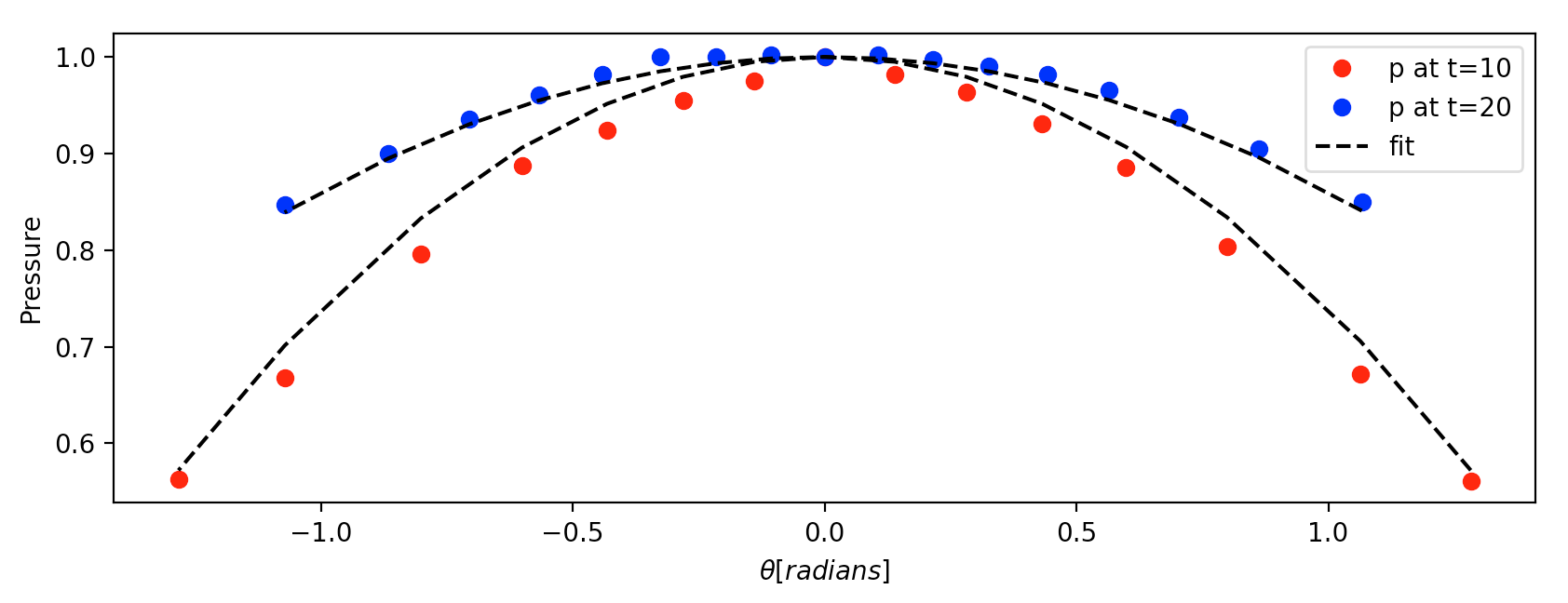}
 %  \caption{}
   \label{fig:p_ang}
%\end{subfigure}
%\begin{subfigure}[b]{0.475\textwidth}
   \includegraphics[width=18cm]{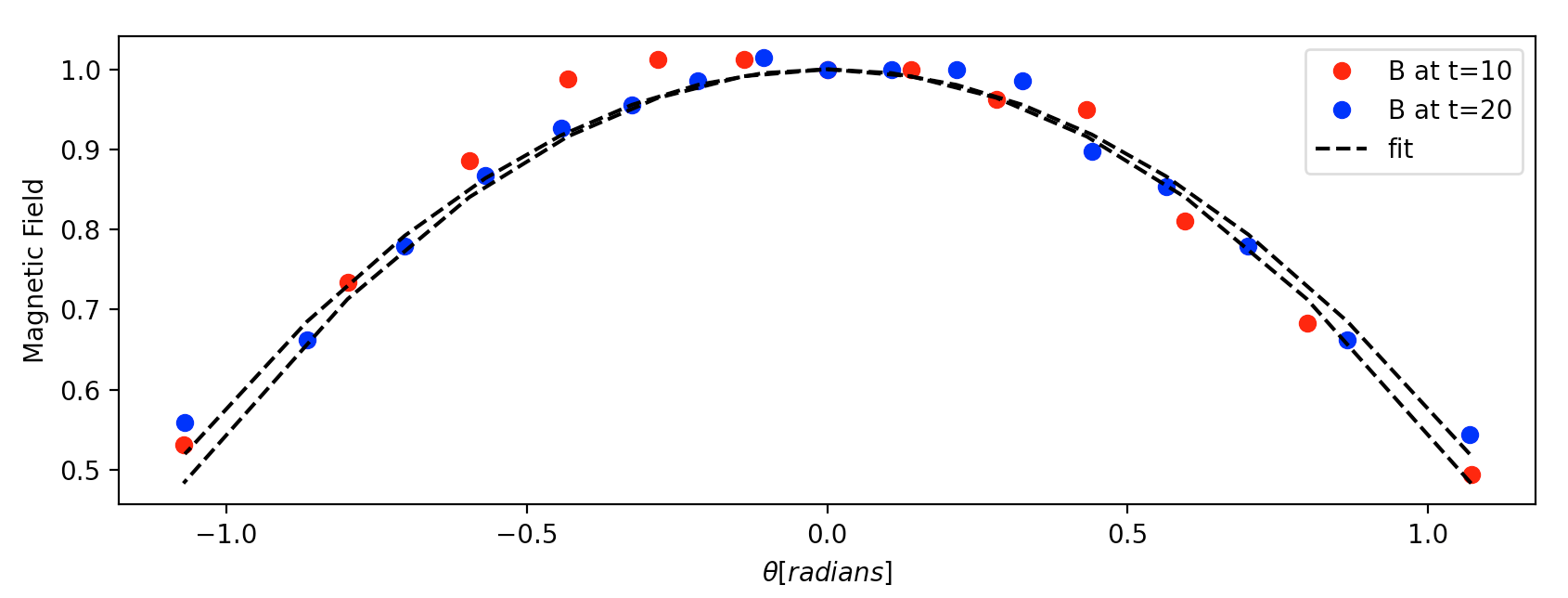}
  %  \caption{}
    \label{fig:B_ang}
%\end{subfigure}
 \caption{Angular variation of pressure and magnetic field at $t=10$ (red circles) and 20 (blue circles) to describe self-similarity of the forward shock. $\theta$ is the angle between the $x$-axis and the line joining the center of the shock and the point of peak pressure/magnetic field along the FS when seen along several $z=0$ cuts in the $xz$ plane. (a) The fitting equations of pressure are $p(\theta)_{t=10} = 1 - 0.27\theta^2$ and $p(\theta)_{t=20} = 1 - 0.14\theta^2$. (b) The fitting equations of magnetic field are $B(\theta)_{t=10} = 1 - 0.45\theta^2$ and $B(\theta)_{t=20} = 1 - 0.42\theta^2$. The pressure profile does not retain a self-similar shape over time as discussed in section \ref{sec:ZR_DIS}.}
\label{fig:p_B_ang}
\end{figure*}

\begin{figure*}	
   \includegraphics[width=18cm]{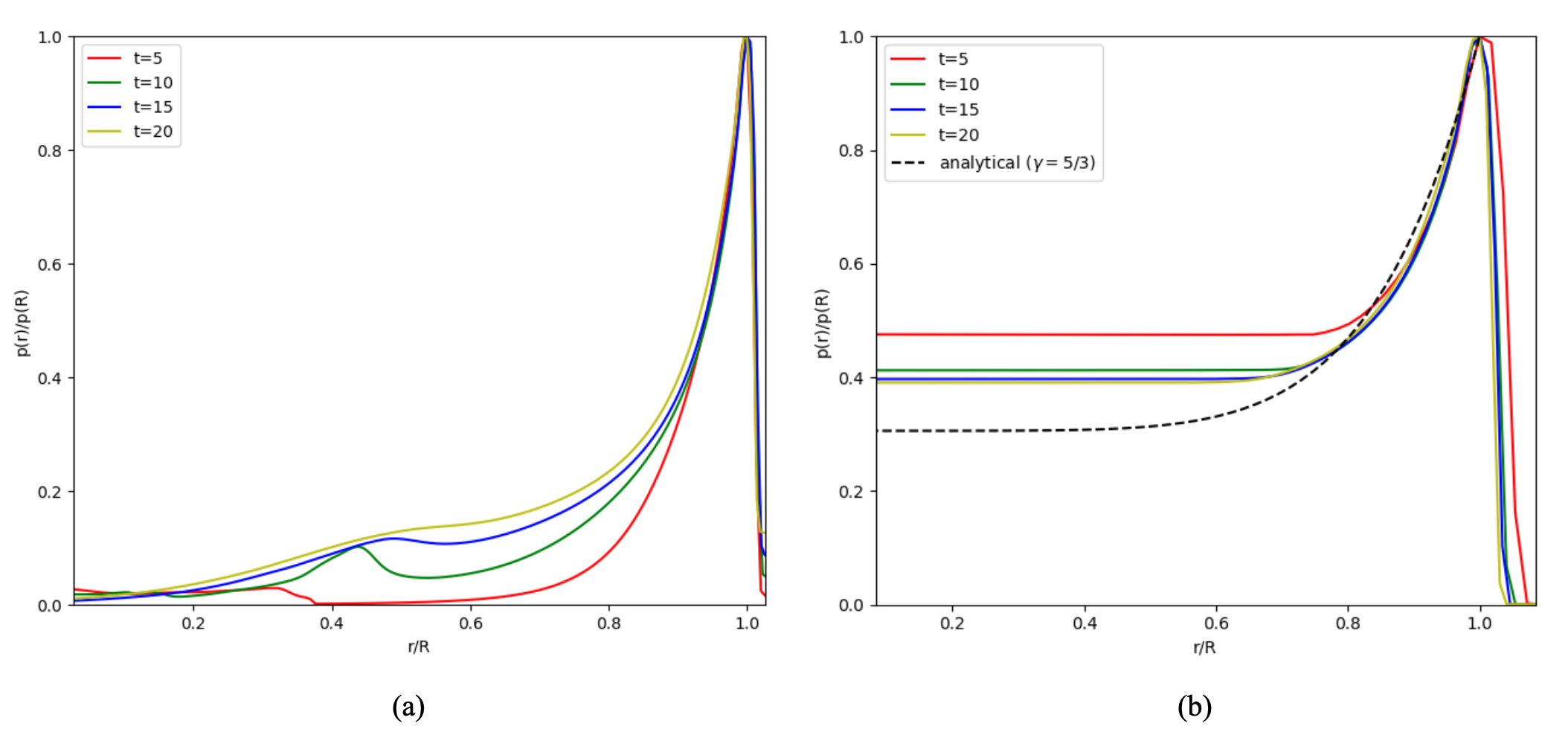}
    \caption{Internal structure of the shocked ISM compared to the Sedov-Taylor case. Plotted are normalized pressure versus radius of the FS of the ``Impact'' case (panel a) and Sedov-Taylor (panel b). In the impact case, the  pressure  in the bulk of the flow drops rapidly due to back flow of material and reaches a much lower limiting value than that of a Sedov-Taylor blastwave.}
    \label{fig:impact&sedov}
\end{figure*}

Below we show that although the radial dynamics of the FS is approximately self-similar (see figure \ref{fig:p_R_TE}), the lateral and internal structures are not.

Using the pressure and magnetic field maps (figures~\ref{fig:prs_map} and~\ref{fig:b3d_map}), we  plot the angular variation of pressure and magnetic field of the FS at $t=10$ and 20 as shown in figure~\ref{fig:p_B_ang}. $\theta$ is the angle in radians, between the $x$ axis and the line joining the center of the shock and the point of peak pressure/magnetic field along the FS when seen along several $z=0$ cuts in the $xz$ plane. Because the ISM-cavity interface moves with time, the shock center shifts slightly at the two time instances which has been accounted for while plotting the angular variation. We quantify the shape of the shock  by fitting a polynomial to the numerical pressure and magnetic field data as shown by black dashed lines. %The functional forms of pressure and magnetic field, thus obtained at $t=10$ and $t=20$ are $p(\theta)_{t=10} = 11.4(1 - 0.27\theta^2)$, $p(\theta)_{t=20} = 4.34(1 - 0.13\theta^2)$, $B(\theta)_{t=10} = 0.8(1 - 0.45\theta^2)$ and $B(\theta)_{t=20} = 0.68(1 - 0.41\theta^2)$. 
As seen, pressure of the FS does not retain a self-similar shape over time in line with the discussion of section \ref{sec:ZR_DIS}.

The lateral dependance of \Bf\ is consistent with $\propto \cos\theta$ scaling, as expected from a point explosion in constant \Bf. On the other hand,  the scaling of pressure evolves with time, becoming flatter (more spherically symmetric). This indicates that the structure of the shocked medium with time evolves towards becoming more spherical, Sedov-Taylor-like solution. 

Another measure to demonstrate the FS's deviation from self-similarity is to compare its pressure (in units of the immediate post-shock value) versus radius (normalized by the shock radius) to the numerical and analytical Sedov solutions for a spherical blastwave as shown in figure~\ref{fig:impact&sedov}. To make this comparison, we run low-resolution simulations ($N_{\rm X}=N_{\rm Y}=N_{\rm Z}$ = 256) of the radio blob and spherical Sedov blastwave. For brevity, we call the former ``Impact'' case and the latter ``Sedov'' case. We compare the radius dependence of pressure using the pressure map similar to figure~\ref{fig:prs_map} to the numerical Sedov solution as well as the analytical Sedov solution reproduced from \citet{Shu1992} (figure 17.3) at $t=5, 10, 15$ and 20. In case of the analytical Sedov solution (dashed black line in panel (b) of figure \ref{fig:impact&sedov}), pressure $p$ reaches a limiting value of 0.306 (for $\gamma = 5/3$) as $r/R \rightarrow 0$. Numerical Sedov blastwave solutions approach this analytical value at later times as expected. For the impact case, due to the ``exhaust'' as discussed in section~\ref{sec:results}, pressure drops more rapidly and approaches a much lower limiting value ($\sim 0$) than the Sedov limiting value. This again validates the discussion of section \ref{sec:ZR_DIS} that overall dynamics is not self-similar.

Finally, the intermediate self-similarity of FS between momentum and energy conserving regimes is established according to the discussion of section~\ref{sec:ZR_DIS} where $n \approx 1.05$ - an excellent agreement between theory and numerical results.

\subsection{Effects of  reconnection between ejecta and ISM}
\label{sec:recon_results}

\begin{figure*}	
   \includegraphics[width=18cm]{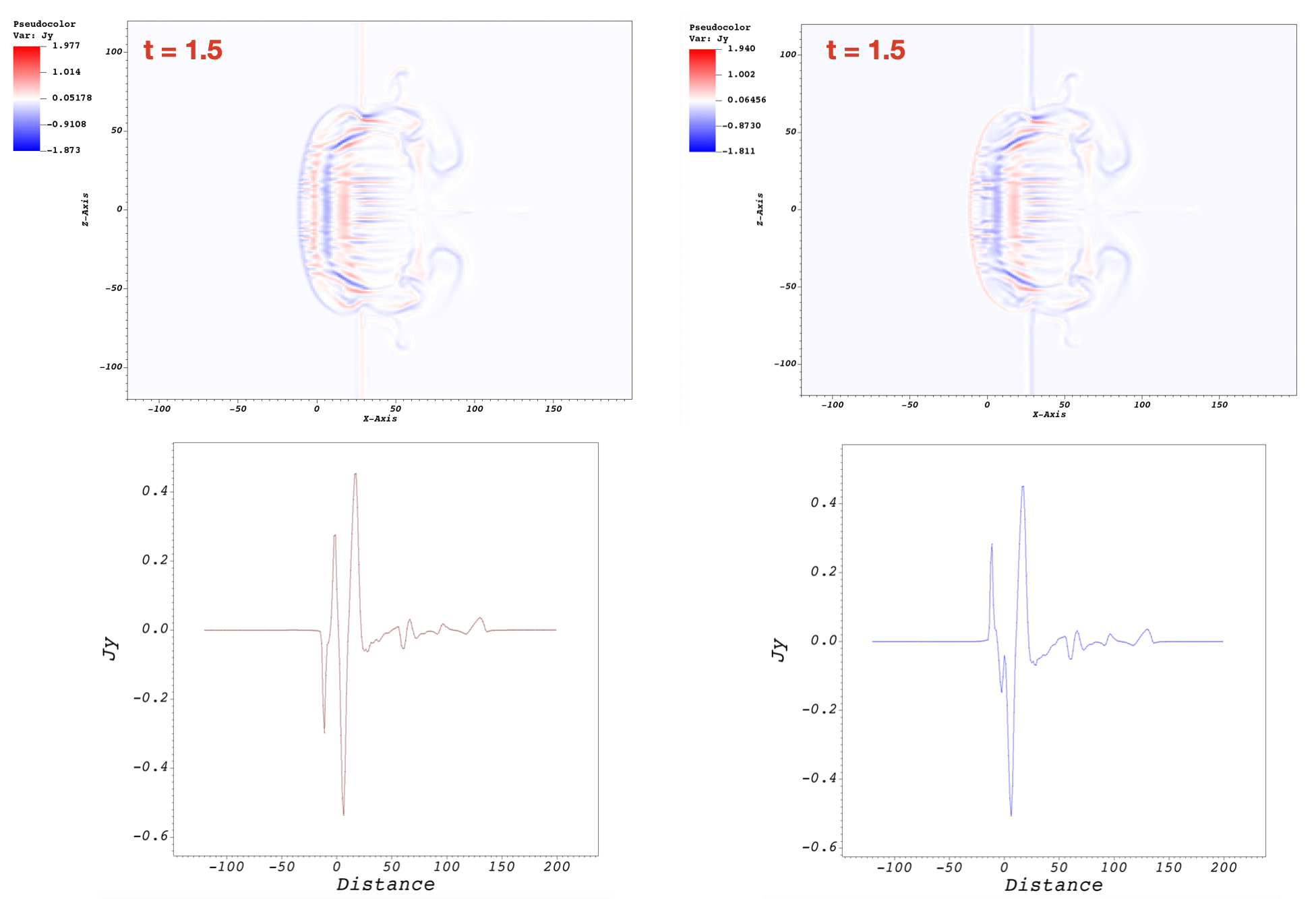}
    \caption{Analysis of magnetic reconnection in terms of currents plotted in 2-D and 1-D for the parallel and anti-parallel set-ups. Upper two panels going from left to right are slices in the $xz$ plane of the low-resolution MHD simulation of the interaction of radio blob with a parallel and anti-parallel ISM magnetic field, respectively, captured at $t=1.5$. Colors indicate $y$ component of current $J_y$. Lower two panels are $Jy$ profiles plotted along $z=0$ at the same time instant. For the parallel case (left panels), magnetic field compression at the forward shock is followed by a strong current sheet followed by current in blob material. For the anti-parallel case (right panels), field compression at the FS is identical to the parallel case but with a reversed orientation, whereas the current sheet following it is much weaker than the parallel case.}
    \label{fig:Jy_2D}
\end{figure*}

\begin{figure*}	
   \includegraphics[width=18cm]{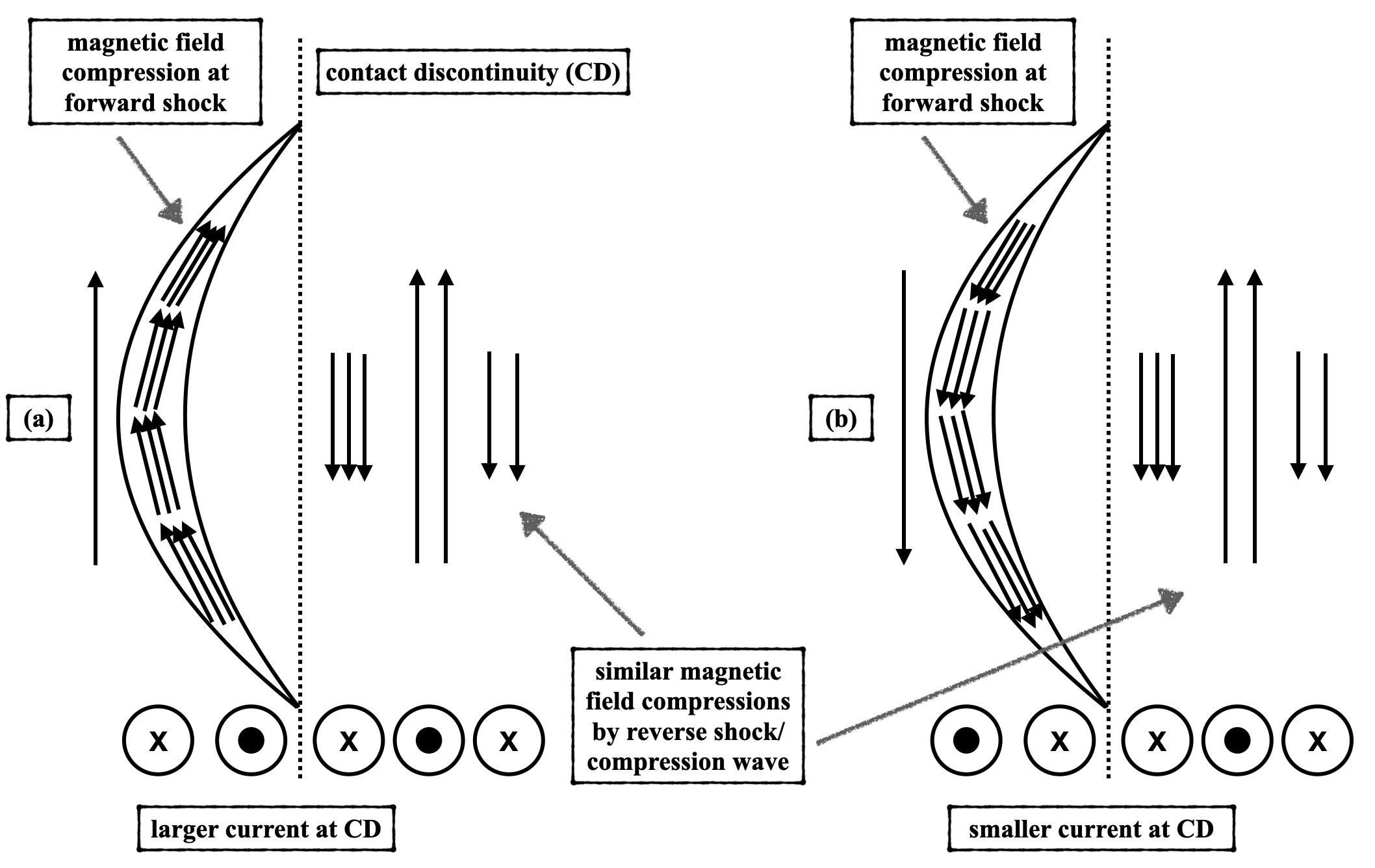}
    \caption{Qualitative picture of magnetic field compression at the forward shock and current sheet formation at the contact discontinuity followed by similar field compressions in the blob by the reverse shock/compression wave in the (a) parallel and (b) anti-parallel orientations. Arrows depict magnetic field vectors in the $xz$ plane and current $J_y$ is shown going into or out of the page.}
    \label{fig:currents}
\end{figure*}

Various orientations of the \Bf\ of the blob as shown in panel \ref{fig:setup} are expected to result in different magnetic configurations at early times, figure~\ref{fig:Jy_2D}: some will be more prone to reconnection. We investigate these effects next. Upper panels of figure \ref{fig:Jy_2D} depict the $y$ component of current $J_y$, where $\mathbfit{J} = \nabla \times \mathbfit{B}$ for the parallel and anti-parallel cases, respectively, projected in 2-D ($xz$ plane) at $t=1.5$. Lower panels are profiles of $J_y$ plotted along $z=0$ at the same time instant. Figure \ref{fig:currents} is a schematic to understand current formation and magnetic reconnection. %{\color{red} The Jy has a striped structure, especially at t=3, I doubt in the meaning of 1D plots... Pink color at right top panel???}

At $t=1.5$, going from left to right, magnetic field compression at the forward shock (shown by congregated arrows in figure \ref{fig:currents}) is accompanied by equal magnitudes of current for parallel and anti-parallel cases. However, for the parallel case, field compression is followed by a strong current sheet near the contact discontinuity (CD) as seen from the relative magnitudes of $J_y$ for the two cases at a distance of $\sim -10$. These current peaks are then followed by current in the blob material, evolving identically in both cases. In other words, both parallel and antiparallel cases  exhibit similar magnetic field compressions by the reverse shock passing through the blob material. Thus, numerical resistivity is not significant even in the highly compressed regions with oppositely directed \Bfs.

 %At $t=3$, the same pattern follows except that the field compressions at the FS get significantly weaker for both cases compared to their earlier time counterparts, whereas current sheet at the CD becomes stronger. 

Our results do not show any significant effects of magnetic reconnection as we do not see a considerable difference in the magnetic field structure between the parallel and anti-parallel orientations. Magnetic reconnection at the CD in the parallel case plays only a mild role and field evolution is mostly dominated by MHD-type dynamics. For the analysis of shock properties and describing the radio emission, we perform a high-resolution simulation of the blob-ISM interaction with the anti-parallel orientation to minimize the effects of magnetic reconnection, albeit a weak influence (see section \ref{sec:num_set} and panel \ref{fig:setup}).

\section{Emissivity maps and light curves}
\label{sec:emiss_maps}

\subsection{Synthetic synchrotron emissivity}

To analyze the radio emission from the 2004 GF, we create integrated synthetic synchrotron emissivity maps using models of \citet{Chevalier1982} and \citet{Chevalier1996} which estimate the evolution of synchrotron radio emission from supernova remnants (SNRs). %described as Sedov remnants. 
Recall that our 3-D numerical results of pressure and radius are qualitatively similar to the evolution of a Sedov-Taylor blastwave, namely that the hydrodynamic evolution is approximately  self-similar and $R \propto t^m$.

If absorption effects are neglected, the synchrotron luminosity of a radio SNR is \citep{Chevalier1982} 
\begin{equation}
    L_{\nu} \propto 4\pi R^2 \Delta R K B^{\frac{\gamma +1}{2}} \nu^{-\frac{\gamma - 1}{2}}
    \label{lum}
\end{equation}
Thus, the synchrotron volume emissivity is
\begin{equation}
    j_{\nu} \propto K B^{\frac{\gamma +1}{2}}
    \label{emiss}
\end{equation}
The power-law energy distribution of emitting electrons is $N(E)=KE^{-\gamma}$, where $E$ is energy and $K$ is a constant. The value of $\gamma$ is determined by the observed  radio spectrum. Observations of radio emission of the 2004 GF report the emission spectrum to be characterized by an unbroken power law with $\gamma=2.5$ \citep{Gaensler2005}.

We create two emissivity maps: map1  (amplification of the field at the shock),   and map2 (pure compression of \Bf), see  \citet{Reynolds2017}. Turbulent magnetic field amplification is a prominent mechanism for synchrotron emission in SNRs. \cite{Reynolds1981, Duric1986, Huang1994} discuss that to account for the observed radio synchrotron emission on average, magnetic field inside SNRs must be much higher than typical ISM values, $B_\text{ISM} \approx 5\mu$G. A possible mechanism of magnetic field amplification in SNRs considered by \cite{Gull1973,1983IAUS..101..183F, Duric1986} is that of Rayleigh-Taylor instability of the CD, separating ejecta from the swept-up ISM. On the other hand, a very efficiently accelerated nuclear cosmic ray (CR) component can cause nonlinear magnetic field amplification near the shock \citep{Lucek2000}. Observational evidence of such a mechanism is presented by analyzing the synchrotron spectrum of SN 1006 \citep{Berezhko2002}, Cassiopeia A \citep{Berezhko2004} and Tycho's SNR \citep{Volk2005}. In addition, \cite{Kosenko2011,Kosenko2014} present numerical models to describe the role of cosmic ray acceleration in the evolution of SNRs, with particular focus on Tycho and SN 1006.

Map1 is based on the assumption of magnetic field amplification where the shock puts constant fractions of post-shock pressure $p=\rho_0 V_s^2$ ($\rho_0$ is the ISM density, $V_s$ is the shock speed) into the magnetic field energy \citep{Bell2001} and electron energy. Thus, $K\propto p$ and $B^2\propto p$ giving $j_{\nu} \propto p^{\frac{\gamma+5}{4}}$. Map2 is based on the assumption that instead of being amplified, magnetic field is simply compressed from a uniform upstream value. In this case, $j_{\nu} \propto pB^{\frac{\gamma+1}{2}}$. With $\gamma=2.5$, the two synthetic synchrotron maps are defined by the following scalings
\begin{equation}
    j_{\nu,\text{map1}} \propto p^{1.875}
    \label{map1}
\end{equation}
\begin{equation}
    j_{\nu,\text{map2}} \propto pB^{1.75}
    \label{map2}
\end{equation}

\subsection{Emissivity maps}
\label{EMISS_RESULTS}

\begin{figure*}	
   \includegraphics[width=18cm]{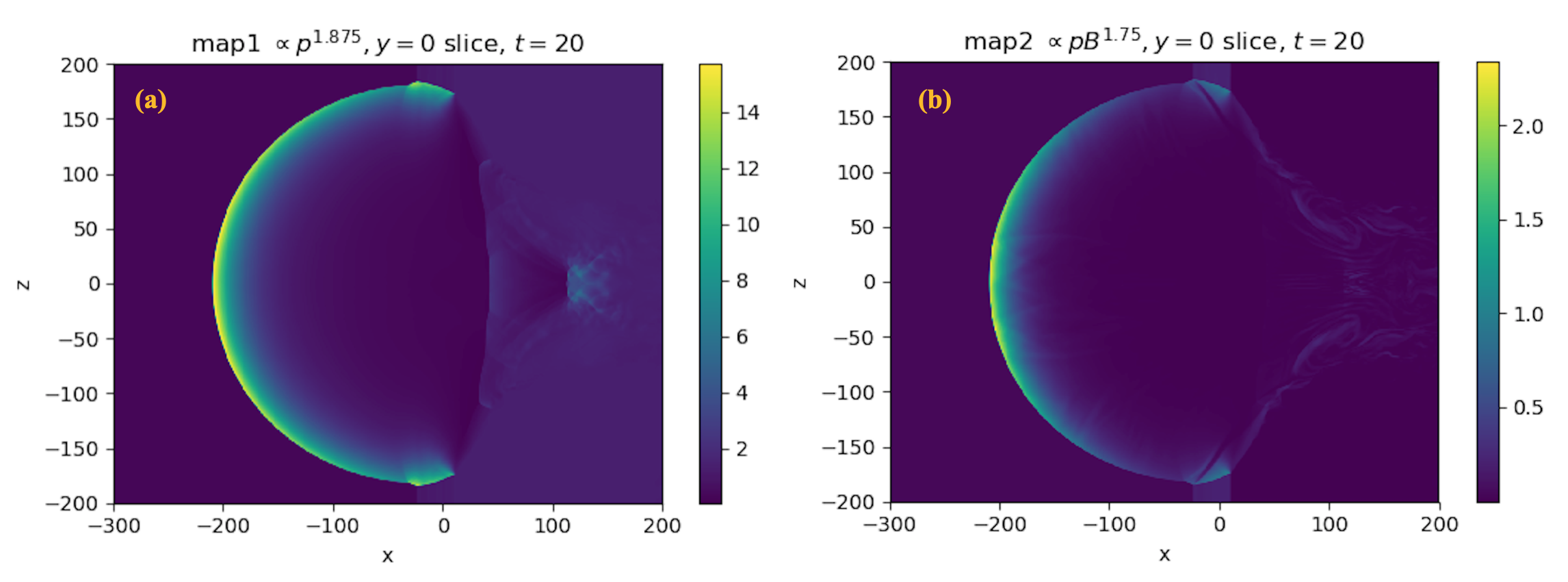}
    \caption{Slices in the xz plane ($y=0$) depicting synthetic synchrotron emissivity at late times, $t=20$, created from the high-resolution MHD simulation of the interaction of radio blob with external ISM. Color bars indicate (a) $j_{\nu,\text{map1}} \propto p^{1.875}$ and (b) $j_{\nu,\text{map2}} \propto pB^{1.75}$.}
    \label{fig:emiss_slices}
\end{figure*}

Figure~\ref{fig:emiss_slices} shows 2-D $y=0$ slices of the two synchrotron emissivity maps scaled as $p^{1.875}$ and $pB^{1.75}$ at late times, $t=20$. Panel (a) mimics the pressure map of figure~\ref{fig:prs_map}, just with a different scaling. It is seen that the shock front (apex at $x\sim -200$), which is the site for electron acceleration, is where all the emission originates. A slight rebrightening appearing at $x\sim 110$ is insignificant compared to the emission at the FS and does not contribute to the temporal evolution of luminosity. FS emission in case of magnetic field compression (panel (b)) is weaker than that with magnetic field amplification, presumably due to dissipation of magnetic field because of the interaction between the magnetized ISM and blob.

\begin{figure*}
   \includegraphics[width=18cm]{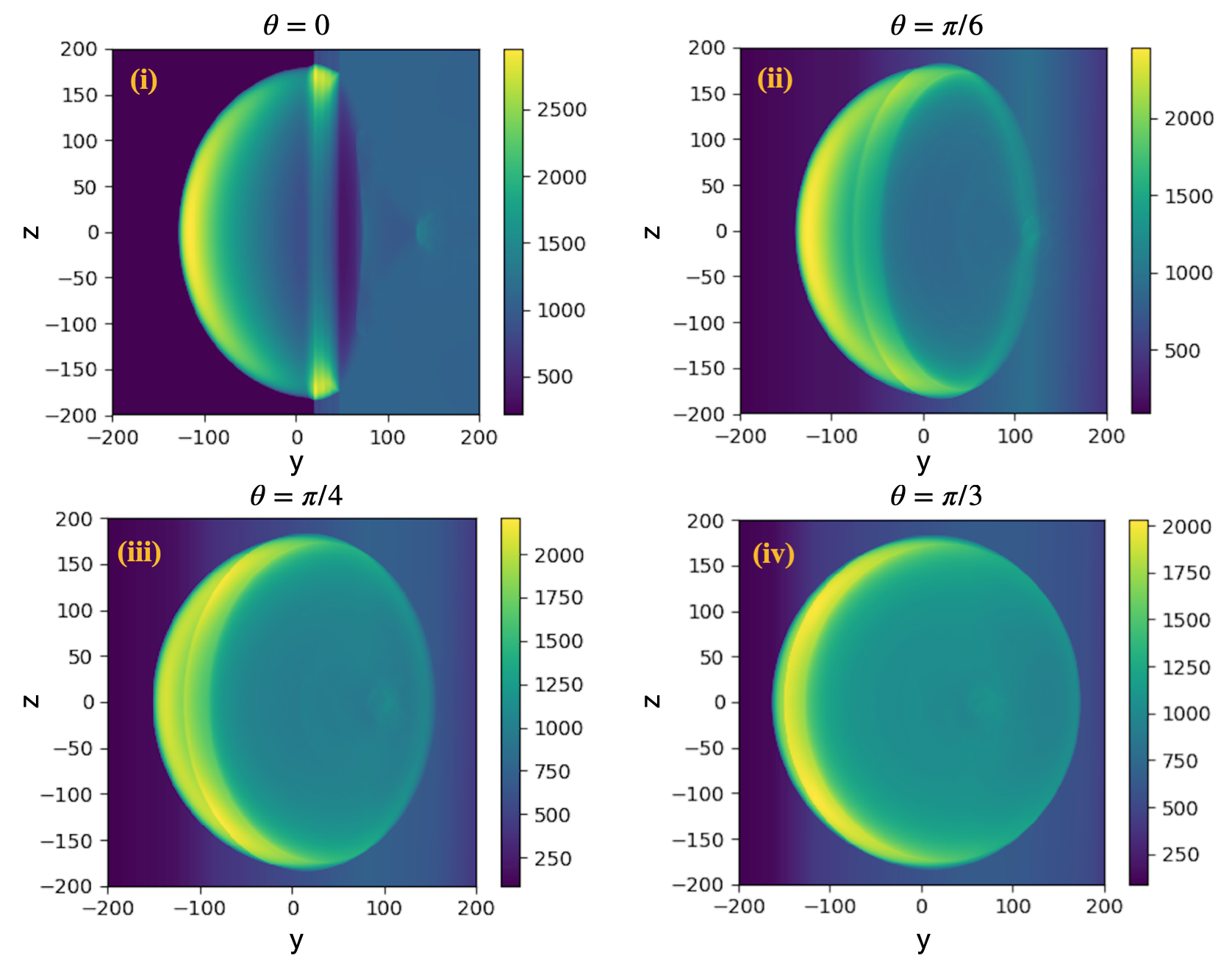}
   \caption{Integrated 2-D emissivity maps for the prescription $p^{1.875}$ - magnetic field amplification, along various lines-of-sight, namely $\theta = 0, \pi/6, \pi/4$ and $\pi/3$, at late times, $t=20$. Emission from the limb dominates the apex emission at all oblique lines-of-sight.}
   \label{fig:j1_int}
\end{figure*}

\begin{figure*}
   \includegraphics[width=18cm]{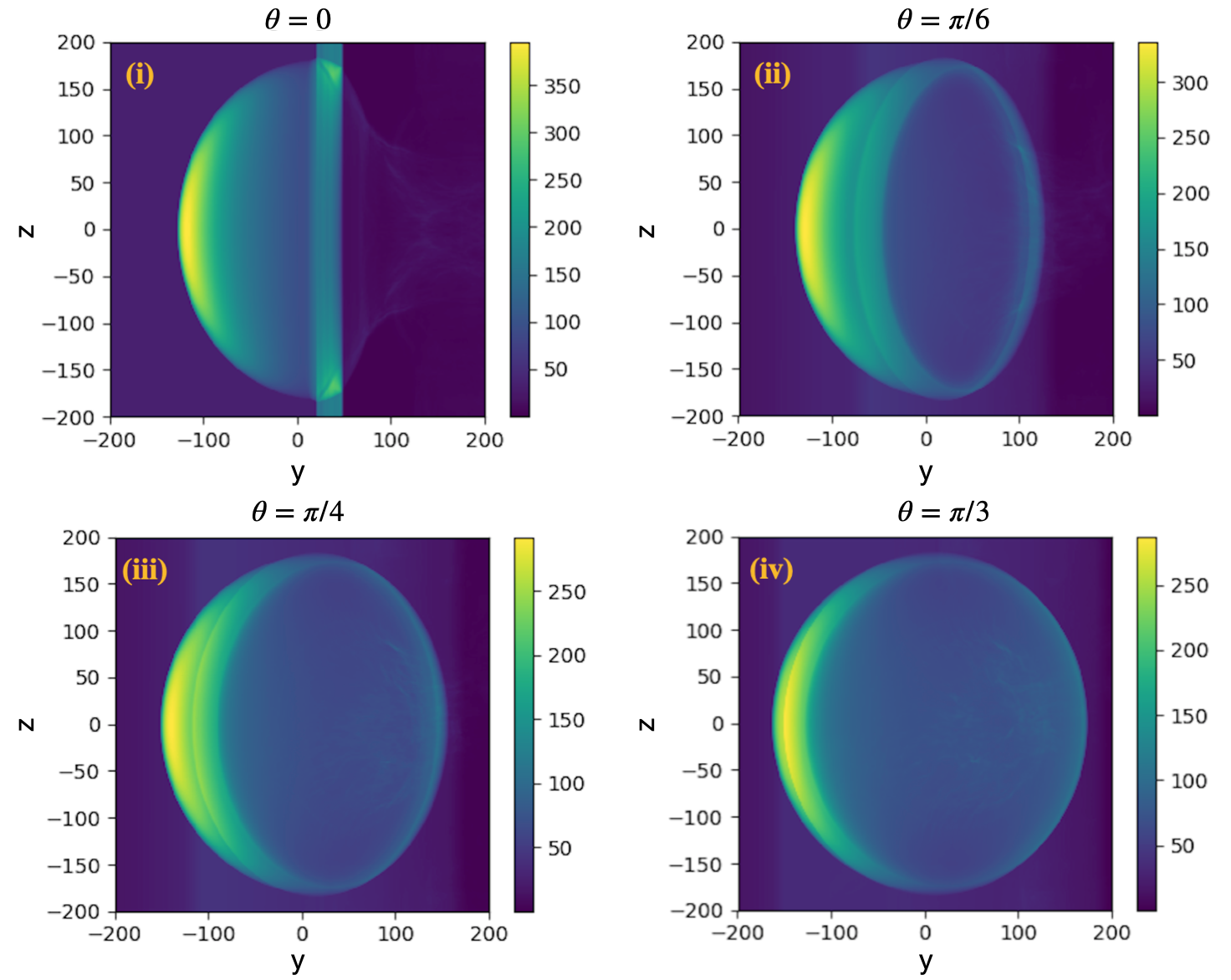}
    \caption{Same as figure \ref{fig:j1_int} but for the prescription $pB^{1.75}$ - magnetic field compression only. Emission is much dimmer than the case with magnetic field amplification.}
    \label{fig:j2_int}
\end{figure*}

Figures \ref{fig:j1_int} and  \ref{fig:j2_int} depict numerical 2-D ($yz$ plane) maps obtained by integrating late-time ($t=20$) synthetic emissivities for the prescriptions $p^{1.875}$ and $pB^{1.75}$, respectively, along lines-of-sight at varying angles from $x$-axis, namely $\theta = 0, \pi/6, \pi/4$ and $\pi/3$. The bright spots near $y=0$ and $z=\pm 150$ occur due to the ISM-wind interaction which has not been subtracted out from the total emission. The case with only magnetic field compression has a weaker integrated emission than the one with magnetic field amplification due to magnetic field interactions. It is seen from both maps that emission from the limb (outer boundary of the spherical emission) is bright and dominates over the apex ($z=0$) emission along all oblique ($\theta>0$) lines-of-sight.

%%%%%%%%%%%%%%%%%%%%%%%%%%%%%%%%%%%%%%%%%

\begin{figure*}	
\centering
   \includegraphics[width=18cm]{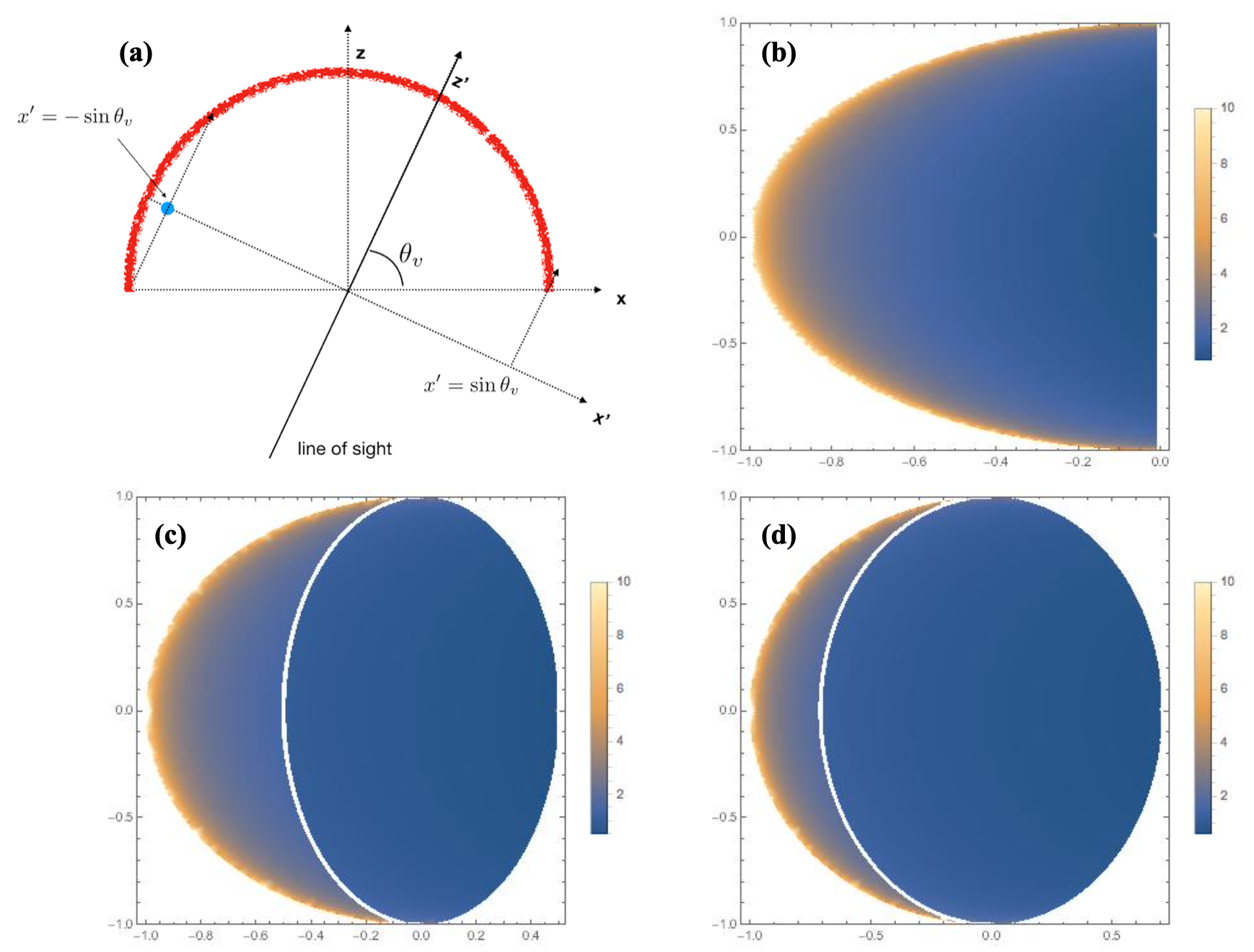}
 \caption{Towards an analytical model of emissivity. (a) Illustration of the model: an  emitting semi-sphere is viewed at angle $\theta_v$. (b) $\theta_v=0$, (c) $\theta_v=\pi/6$, and (d) $ \theta_v=\pi/4$. The scales are normalized to the radius of the cavity. The maps are mostly dominated by the line-of-sight effects, not the fact that pressure/emissivity are  higher at the apex point.}
 \label{emiss_ana}
\end{figure*}

In an effort to understand the results of the simulations, we constructed semi-analytical emission maps shown in figure \ref{emiss_ana}. We approximate the emitting volume as a half-sphere $z>0$ with surface brightness given by the normalized pressure fit $p(\theta)_{t=20} = (1 - 0.14\theta^2)$ (at time 20) as described in section \ref{sec:SS}.
The observer is located at angle $\theta_v$. We assume that emission is generated in a thin shell near the surface of the bubble. (As described previously, the post-shock pressure is released through back flow, leading to fast decrease of emissivity). We then calculate the observed emissivity map according to the prescription $p^{1.875}/(\cos \theta_{r-v} + c_1)$, where $\theta_{r-v}$ is the local angle between the line-of-sight and the radial direction in the emitting bubble. Parameter $c_1=0.1$ is introduced to avoid unphysical values of emissivity when the line-of-sight is tangential to the surface.

%\begin{figure}	
%   \includegraphics[width=0.49\columnwidth]{Emission-blob-oblique001a.jpeg}
%     \includegraphics[width=0.49\columnwidth]{Emission-blob-obliquepi2.jpeg}\\
%       \includegraphics[width=0.49\columnwidth]{Emission-blob-obliquepi3.jpeg}
%     \includegraphics[width=0.49\columnwidth]{Emission-blob-obliquepi4.jpeg}
%    \caption{Towards an analytical model of emissivity. Left top panel: Illustration of the model: an  emitting semi-sphere is viewed at angle $\theta_v$.  Top right: 
%$\theta_v=\pi/2$, bottom row: $ \theta_v=\pi/3$ and  $ \theta_v=\pi/4$. The scales are normalized to the radius of the cavity. The maps are mostly dominated by the  line-of-sight effects, not the fact that  pressure/emissivity are  higher at the apex point.}
%    \label{Emission-blob-oblique}
%\end{figure}

Our semi-analytical results reproduce qualitatively the numerics of figures \ref{fig:j1_int} and \ref{fig:j2_int}. Emissivity shows strong limb brightening due to longer path through emitting volume. On the other hand, effects of pressure (and thus emissivity) variations  along the shock produce only  mild variations of intensity.

\subsection{Light curves}

\begin{figure}	
   \includegraphics[width=\columnwidth]{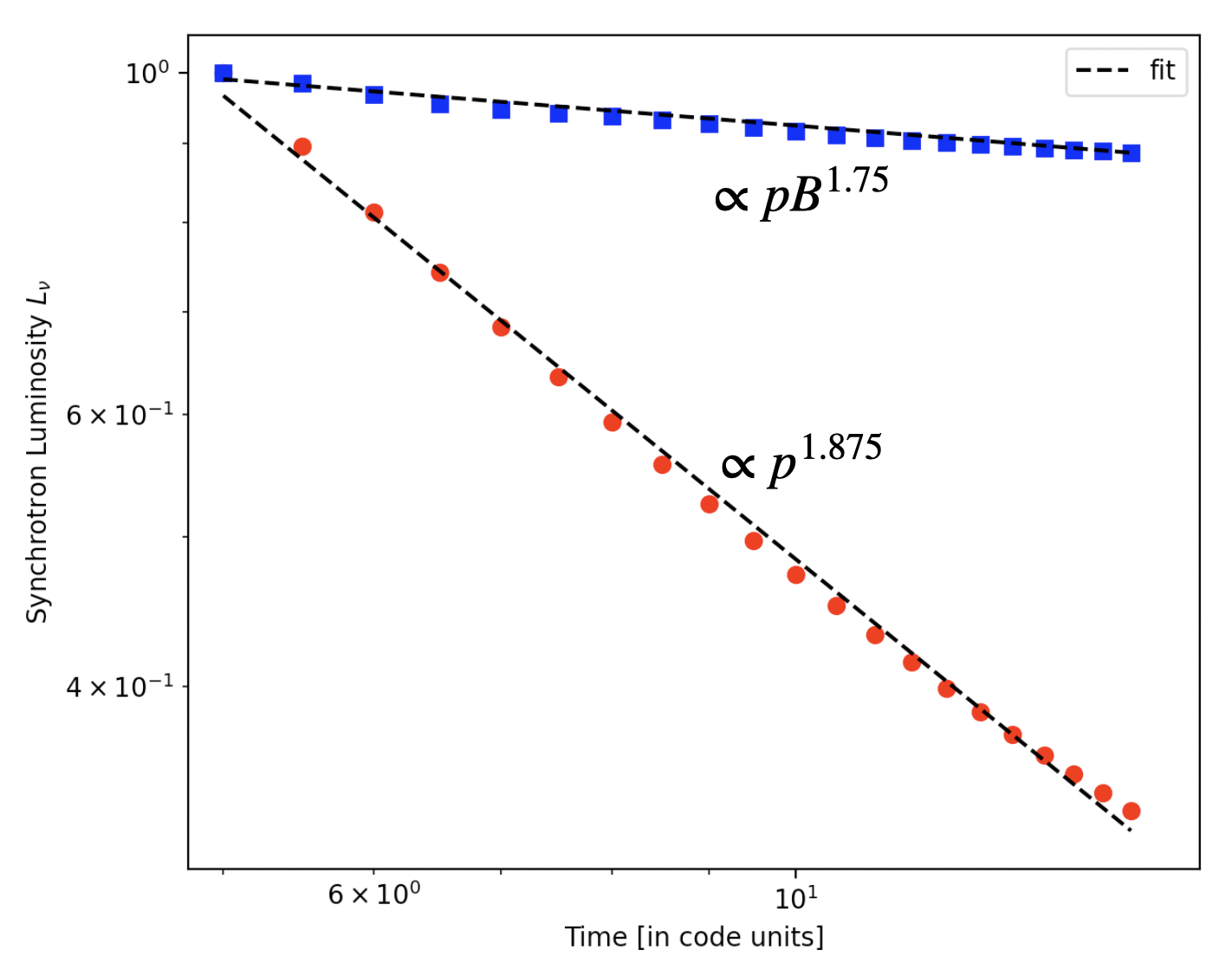}
    \caption{Temporal evolution of normalized synchrotron luminosity of the shocked ISM. Fits through the linear part of the light curves (black dashed lines) give $L_{\nu,\text{map1}} \propto t^{-1}$ and $L_{\nu,\text{map2}} \propto t^{-0.1}$. }
    \label{fig:emiss_TE}
\end{figure}

In figure~\ref{fig:emiss_TE}, we show the temporal evolution of normalized luminosities obtained by integrating  emissivity assuming an optically thin medium and isotropic emission, following the two scalings described by~\ref{map1} and~\ref{map2} over the entire computational box. Synthetic light curves are plotted from $t=5$ to $t=15$ in steps of 0.5 in log-log scale for a lower resolution $N_{\rm X}=N_{\rm Y}=N_{\rm Z}$ = 256. Unlike the integrated 2-D maps of figures \ref{fig:j1_int} and \ref{fig:j2_int}, emission from the ``exhaust flow'' due to the ISM-wind interaction has been subtracted from the total emission to account only for the emission from the shocked ISM. To do this, we employed two tracers, one to trace the ISM material and another to trace the wind and blob. These tracers are ``passive'', meaning they obey simple advection equations and behave as ``tags'' attached to the ISM and blob+wind so that the contributions from each can be isolated. A linear fit to the plots gives power laws for synchrotron luminosities: $L_{\nu,\text{map1}} \propto t^{-1}$ and $L_{\nu,\text{map2}} \propto t^{-0.1}$. As reported in observations, the radio light curve evolves as $\sim t^{-1}$ after undergoing rebrightening, which marks the self-similar Sedov-Taylor evolution phase of the radio nebula. Our synthetic emissivity map based on the assumption of magnetic field amplification is able to account for this observed temporal evolution. In addition, temporal evolution of numerical emissivity closely matches that of an ideal Sedov remnant with magnetic field amplification, characterized by $R(t) \propto t^{m}$ and $m=0.4$. For such a case, the shock puts constant fractions of the post-shock pressure into electron energy and magnetic-field energy, that is, $K \propto \rho_0 V_s^2$ and $B^2 \propto \rho_0 V_s^2$. Since volume integrated emissivity $L_{\nu} \propto R^3KB^{1+\alpha}$ and $V_s \propto t^{m-1}$, $L_{\nu} \propto t^{m(6+\alpha) - (3+\alpha)}$ \citep{Reynolds2017}, where $\alpha = \frac{\gamma - 1}{2} = 0.75$ based on observations. Thus, $L_{\nu} \propto t^{-1.05}$. %This discrepancy between numerics and an ideal Sedov remnant is expected because of the faster pressure drop due to considerable ``exhaust flow'' in the impact case. 
On the other hand, if we use the values of exponents from the simulated radius and pressure evolution, that is, $R(t) \propto t^{m_1}$ and $p(t) \propto t^{m_2}$ (section \ref{sec:p_v_R}), unlike the ideal Sedov case where pressure and radius are mutually dependent, volume integrated emissivity $L_{\nu} \propto R^3p^{(3+\alpha)/2} \propto t^{[6m_1 + m_2(3+\alpha)]/2}$. This gives a semi-analytical estimate of luminosity $L_{\nu} \propto t^{-1.4}$, consistent with numerics. The slight discrepancy is expected because of the faster pressure drop due to considerable ``exhaust flow'' in the impact case. It is also important to note that the nearly flat light curve in case of field compression $L_{\nu,\text{map2}} \propto t^{-0.1}$ (blue squares) is a consequence of energy conservation and would be exactly constant without a significant backflow.

%To compare with observations of radio emission from SNRs, we define a secular decline rate $-d = j_{\nu}^{-1}dj_{\nu}/dt$. \cite{Vinyajkin1999} report such a decline rate of $0.92 \text{yr}^{-1}$ for Tycho Brahe's SNR radio emission which is frequency-independent. Within the frequency range of 86-5000 MHz, the weighted mean value of the decline rate is $0.41 \text{yr}^{-1}$. A list of values for the decline rate for the radio emission from CasA at different frequencies is tabulated in \cite{Vinyajkin2014}. The secular decline rate has values ranging from $0.3 - 1 \text{yr}^{-1}$. Another estimate of Tycho's flux decrease rate is quoted by \cite{Stankevich2003} who report that the luminosity at 1667 MHz is weakening with an annual mean rate of $0.47\%$. A corresponding rate for Kepler's SNR is $0.2\%$. Lastly, \cite{Welier2010} present the evolution of secular decrease rate from their extensive observations of the radio emission from SN 1993J in the frequency range 85-110 GHz. The flux decline rate steepens from $\sim 0.7$ to $\sim -2.7$ at $\sim 3100$ days after the shock breakout, which they describe as an exponential decay, rather than a power-law decline. From MHD simulations, we get $-d = 0.7$ (per unit time) for the case with magnetic field amplification and $-d = 0.1$ (per unit time) for the case without amplification. Considering the simplicity of our MHD model and overall uncertainties and errors reported in observations, these secular decline rates are reasonably consistent with those of popular radio SNRs. 

%\begin{figure}	
%   \includegraphics[width=\columnwidth]{emiss_m1_map.png}
%    \caption{Same as figure~\ref{fig:prs_map} but colors indicate map1 of synchrotron emissivity defined as $\propto p^{1.875}$.}
%    \label{fig:m1_map}
%\end{figure}
%
%\begin{figure}	
%   \includegraphics[width=\columnwidth]{emiss_m2_map.png}
%    \caption{Same as figure~\ref{fig:prs_map} but colors indicate map2 of synchrotron emissivity defined as $\propto pB^{1.75}$.}
%    \label{fig:m2_map}
%\end{figure}

\section{Discussion and conclusions}
\label{sec:DIS}

In this work we model the radio afterglow emission   from the 27 December 2004 giant flare from SGR 1806-20. 
We consider the  interaction of light, magnetically dominated cloud  (CME)  ejected during the magnetar flare with the surrounding ISM. We identify the observed emission features with the forward shock created by the impact.  The  magnetic blob is first advected with the magnetar wind, and later impacts on the ISM. The impact creates  a forward shock, and a complicated backward flow.  

Since the CME implants both energy and momentum into the ISM, the flow is not self-similar. Yet, we find that the dynamics of the forward shock (\textit{e.g.} motion of the apex point) closely follows the Sedov-Taylor (energy conserving) prescription. We find that the lateral structure of the shock is not self-similar, it evolves with time. At late times, it approaches Sedov-Taylor like spherical symmetry. At the same time, internal structure of the shocked material is strongly modified by the backward exhaust flow: the pressure in the bulk  decreases much faster than in Sedov-Taylor case.

We identify a number of magnetohydrodynamic features of the interaction that can contribute to   particle acceleration and  the production of radio emission: (i)  reconnection between the CME's  and ISM's \Bf; (ii) forward shock generated in the ISM; (iii) reverse shock in the CME, and (iv) shocks in the exhaust flow of the ISM. 

Relative importance of these contributions is expected to depend mostly on the parameters of  the CME. We adopt a magnetically-dominated paradigm for the ejected CME (if a flare was a magnetospheric event, large baryonic loading is not expected).  The CME is then ``light  and fast'': it carries a lot of energy, but not much mass or momentum. As a result of this assumption, the contribution from the reverse shock in the ejecta  is insignificant.  
We also find that for our parameters, reconnection between the internal and ISM \Bfs\ does not contribute considerably: it is the forward shock and the shocks in the exhaust flow that dominate the pressure. 

To compare with observations we employ two prescriptions to connect MHD properties with radio emissivity: pure compression of the \Bf, as well as turbulent amplification. This is, naturally, the weakest point of the model. Accounting for radio emission, that carries a tiny fraction of luminosity  and total energetics, is a notoriously difficult problem in the study of SNRs \citep{Kennel1984,Atoyan1999,Reynolds2017}. However, the mechanism of magnetic field amplification is often invoked to explain synchrotron emission in SNRs \citep{Reynolds1981, Duric1986, Huang1994}.

Observations \citep{Cameron2005,Gaensler2005} are similarly complicated, showing somewhat different evolution at different frequencies and changing temporal behavior. However, considering the simplicity of the MHD model and complications involved in quantifying radio synchrotron emission, our analysis with field amplification produces a light curve ($\sim t^{-1}$) consistent with the observed temporal evolution of the SGR 1806-20 radio nebula ($\sim t^{-1.1}$) in the self-similar phase. This, in turn, could mean that field amplification might indeed play a significant role in radio afterglow from magnetar giant flares.

An interesting feature of the observed radio light curve is rebrightening at $\sim 25$ days when the radio flux increases by a factor of $\sim 2$ \citep{Gelfand2005}. A scenario with a pre-existing shell around the SGR is invoked to explain this flux increase when the emission from swept-up material dominates the light curve \citep{Granot2006, Gelfand2007}. We test this hypothesis through our MHD simulation. Although the presence of a dense shell produces a break in the temporal evolution of flux, we find no evidence of rebrightening in the light curve due to the impact of the magnetic cloud on the shell. Interestingly, the late-time light curve with only magnetic field compression ($\sim t^{-1}$) does better at accounting for observations, if the shell scenario is accepted. 

Several prospects can be explored in the future. One of the limitations of the present  approach is not so small beta-parameter of the blob, $\beta \sim 2$. This is a purely numerical limitation as  the code needs to resolve highly different properties of the wind/the blob and the ISM.

 In this study, interaction of the magnetic blob with a constant-density ISM/shell was considered. It would be worthwhile to explore a broader parameter space for the blob and ISM. For example, interaction between the blob and an ISM with varying density profiles such as a $\rho \propto r^{-2}$ or an exponential profile can be analyzed. In addition, effects of a denser or a larger/smaller blob can be tested. Emissivity maps and light curves can be calculated for both prescriptions, and compared with theoretical estimates for Sedov remnants as discussed by \cite{Reynolds2017} as well as with the observed light curve. This exercise can provide a better understanding about the nature of the outflow ejected by the SGR flare and its astrophysical environment. Secondly, a better investigation of the effect of a pre-existing shell around the SGR can be performed by testing the MHD code in different regimes with multiple combinations of relative shell thickness and density. It is expected that rebrightening might be observed in some regimes over and above the break in the light curve. Lastly, implications of synchrotron self-absorption on the integrated emissivity maps and light curves, an important consideration in radio, should be explored carefully to gain a more accurate understanding of the nature of the radio nebula.

%ML: I STOPPED HERE. I AM A BIT CONFUSED BY THE PREDICTED TEMPORAL SCALING FOR DIFFERENT EMISSION PRESCRIPTIONS.
%WE DISCUSSED THIS (PAGE 72 OF THE FILE GEOMETRIES.KEY), BUT I AM MISSING IT HERE. 
%
%ALSO, CAN WE CHANGE THE COLOR SCHEME FOR Figure 11 TO  MORE CONTRAST.

%1. Comparison of plots to Sedov-Taylor: This high-velocity backflow causes the forward shock pressure to fall faster than a Sedov-Taylor blastwave as will be evident from the time evolution plots.
%
%2. Comparison of results to observations of radio emission associated with the SGR 1806-20 giant flare.
%
%3. Rebrightening observed between day 25-33 where flux increase by factor. This is before S-T sets in.
%
%4. After 33 days, flux decreases roughly as t(-1) - emission enters ST phase. B field amplification is able to account for this better.
%
%Our simulations also indicate that
%flux rebrightening reported in observations between 25 and 30 days is unlikely to be caused by   a pre-existing dense ISM shell.

%\section{Conclusions}
%\label{sec:CON}
%
%The last numbered section should briefly summarise what has been done, and describe
%the final conclusions which the authors draw from their work.

\section*{Acknowledgements}

%We thank the anonymous referee for their constructive comments which greatly
%improved the manuscript. 
The numerical simulations were carried out in the CFCA
cluster of National Astronomical Observatory of Japan. We thank the PLUTO team for
the possibility to use the PLUTO code and for technical support. Visualization of results was performed in the VisIt package and using Python. This work had been supported by 
NASA grants 80NSSC17K0757 and 80NSSC20K0910,   NSF grants 1903332 and  1908590.

%%%%%%%%%%%%%%%%%%%%%%%%%%%%%%%%%%%%%%%%%%%%%%%%%%
\section*{Data Availability}

The data underlying this article are available in the article.

% 
%The inclusion of a Data Availability Statement is a requirement for articles published in MNRAS. Data Availability Statements provide a standardised format for readers to understand the availability of data underlying the research results described in the article. The statement may refer to original data generated in the course of the study or to third-party data analysed in the article. The statement should describe and provide means of access, where possible, by linking to the data or providing the required accession numbers for the relevant databases or DOIs.


%\begin{thebibliography}{99}
%\end{thebibliography}

\bibliographystyle{mnras}
\bibliography{riddhi,BibTex-ML} % if your bibtex file is called example.bib

\appendix

\section{Analytical Model: The Grad-Shafranov equation and its solution}
\label{sec:GS} % used for referring to this section from elsewhere

In MHD equilibria, the Lorentz force is balanced by the pressure gradient hence demanding
\begin{equation}
    \nabla p = \mathbfit{J} \times  \mathbfit{B} 
    \label{eq:equilibria}
\end{equation}
where $p$ is plasma pressure, and \mathbfit{J}  and \mathbfit{B} are current density and magnetic field.\\
In the Grad-Shafranov framework \citep{Shafranov1966,Grad1967} the  {\bf axisymmetric} magnetic field can be represented by a scalar flux function $\psi$ in spherical coordinates
\begin{equation}
    \mathbfit{B}=\nabla\psi\times\nabla\phi+\lambda\psi\nabla\phi
    \label{eq:axisB}
\end{equation}
Force balance~\ref{eq:equilibria} gives the Grad-Shafranov equation
\begin{equation}
    \frac{\partial^2\psi}{\partial r^2}+\frac{\sin\theta}{r^2}\frac{\partial}{\partial\theta}\left(\frac{1}{\sin\theta}\frac{\partial\psi}{\partial\theta}\right)+F(\psi)r^2\sin^2\theta + G(\psi)=0
    \label{eq:GSE}
\end{equation}
where $F(\psi) = 4\pi dp/d\psi$.

We model the radio emission using the magnetic field structure of \citet{Gourgouliatos2010} namely a structure demanding vanishing magnetic field on the surface due to unmagnetized external plasma. We will call this the `magnetic blob' henceforth. This mathematical problem requires both Dirichlet and Neumann boundary conditions to be satisfied, meaning both the flux function $\psi$ and its normal derivative $\partial_r\psi$ should be continuous at the boundary. Following \citet{Gourgouliatos2010}, for solutions of~\ref{eq:GSE} of desired properties, $F(\psi) = 4\pi dp/d\psi = F_0$ and $G(\psi) = \lambda^2\psi$ are chosen. This choice along with $\psi = \sin^2\theta f(r)$ leads to the analytical solution for the radial part of $\psi$
\begin{equation}
    f(r)= A_0\lambda r j_1(\lambda r) - \frac{F_0}{\lambda^2} r^2
    \label{eq:radial_f}
\end{equation}
where $j_1(\lambda r) = \frac{\sin(\lambda r)}{\lambda^2 r^2} - \frac{\cos(\lambda r)}{\lambda r}$ is the spherical Bessel function.\\

\noindent{Thus, it is observed that the solution is simply a force-free spheromak superposed with a uniformly twisted magnetic field \citep{Gourgouliatos2010}}
\begin{equation}
    \psi= \sin^2\theta \left(A_0\lambda r j_1(\lambda r) - \frac{F_0}{\lambda^2} r^2\right)
    \label{eq:solution}
\end{equation}

Next, we determine the normalizing constant $A_0$. We consider the flux to be confined within a structure of radius $r_0$ which can take several values unlike \citet{Gourgouliatos2010} where they consider a unit radius. This enables us to write all following equations in terms of $r_0$ giving the flexibility of testing the code for radius dependence, if desired. The first boundary condition of zero magnetic field on the surface $f(r_0) = 0$ gives 
\begin{equation}
    F_0= \frac{A_0\lambda^3}{r_0}j_1(\lambda r_0)
    \label{eq:F_0}
\end{equation}
The second boundary condition of zero surface currents $f'(r_0) = 0$ gives 
\begin{equation}
    \tan(\lambda r_0)= \frac{3\lambda r_0}{3 - \lambda^2 r_0^2}
    \label{eq:tangent}
\end{equation}
The smallest positive root of this equation is $\lambda \approx 5.763/r_0$ giving $F_0 = -31.7A_0/r_0^4$.
Magnetic field components are 
\begin{equation}
    B_r= \frac{2\cos\theta}{r^2}f(r)
    \label{eq:Br}
\end{equation}
\begin{equation}
    B_\theta= -\frac{\sin\theta}{r}f'(r)
    \label{eq:Btheta}
\end{equation}
\begin{equation}
    B_\phi= \frac{\lambda\sin\theta}{r}f(r)
    \label{eq:Bphi}
\end{equation}
Since the blob is held together by magnetic field, pressure profile within the sphere in pressure balance with the ambient medium having pressure $p_0$ can be given as 
\begin{equation}
        p = p_0 + \frac{F_0\psi}{4\pi} = p_0 - \frac{B_\text{max}^2}{8\pi}
    \label{eq:pressure}
\end{equation}
where $B_\text{max}$ is the maximum pressure at the center of the sphere.
Defining a plasma-$\beta$ as $\beta = p_0/B_\text{max}^2$ along with maximizing $f(r)$ and using the value of $F_0$ gives the normalizing constant
\begin{equation}
    A_0 = 0.11\sqrt\frac{p_0}{\beta}r_0^2
    \label{eq:A0}
\end{equation}
As seen from~\ref{eq:pressure}, there is a dip in pressure at the center of the blob. This dip is determined by $B_\text{max}$ and hence $\beta$. To prevent negative pressure at the center, $\beta > 0.5$. Thus, initial magnetic field within the blob used to model the radio emission is completely defined by its initial radius, initial plasma-$\beta$ and pressure of the ambient medium.

\section{Effect of a pre-existing shell around SGR 1806-20}
\label{sec:SHELL}

\begin{figure*}	
\centering
   \includegraphics[width=18cm]{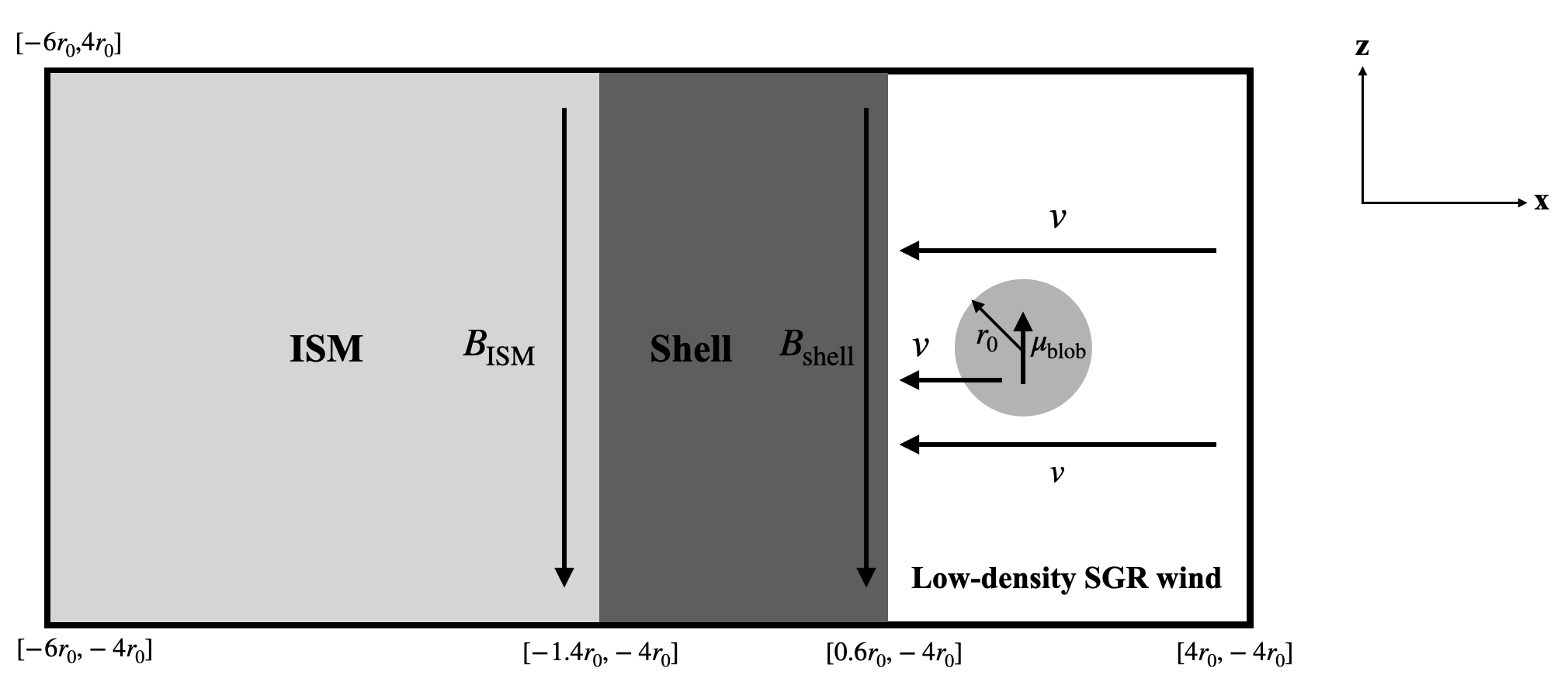}
 \caption{A schematic of the anti-parallel set-up used to assess the effect of a pre-existing shell around the SGR on the shock dynamics. 
 %$v$ is the velocity of the blob and SGR wind moving towards the shell and ISM. Both the shell and ISM magnetic fields are anti-parallel to the blob's magnetic moment to minimize effects of magnetic reconnection. 
 }
 \label{fig:shell_scenario}
\end{figure*}

As a \NS\ moves through the ISM with supersonic velocity, interaction of the wind with the ISM creates a  bow shock \citep{2006ARA&A..44...17G,2019MNRAS.484.4760B}. Post-shock ISM can be approximated as a dense shell.
\cite{Gaensler2005,Granot2006,Gelfand2007} suggested the interaction of the material ejected during the GF with this pre-existing shell at $\sim 10^{16}$ cm as a possible cause for the rebrightening at $\sim 25 $ days.
To test this suggestion we run a long, low-resolution simulation with the shell added to the ISM as shown in figure \ref{fig:shell_scenario}.

\begin{figure*}
   \includegraphics[width=18cm]{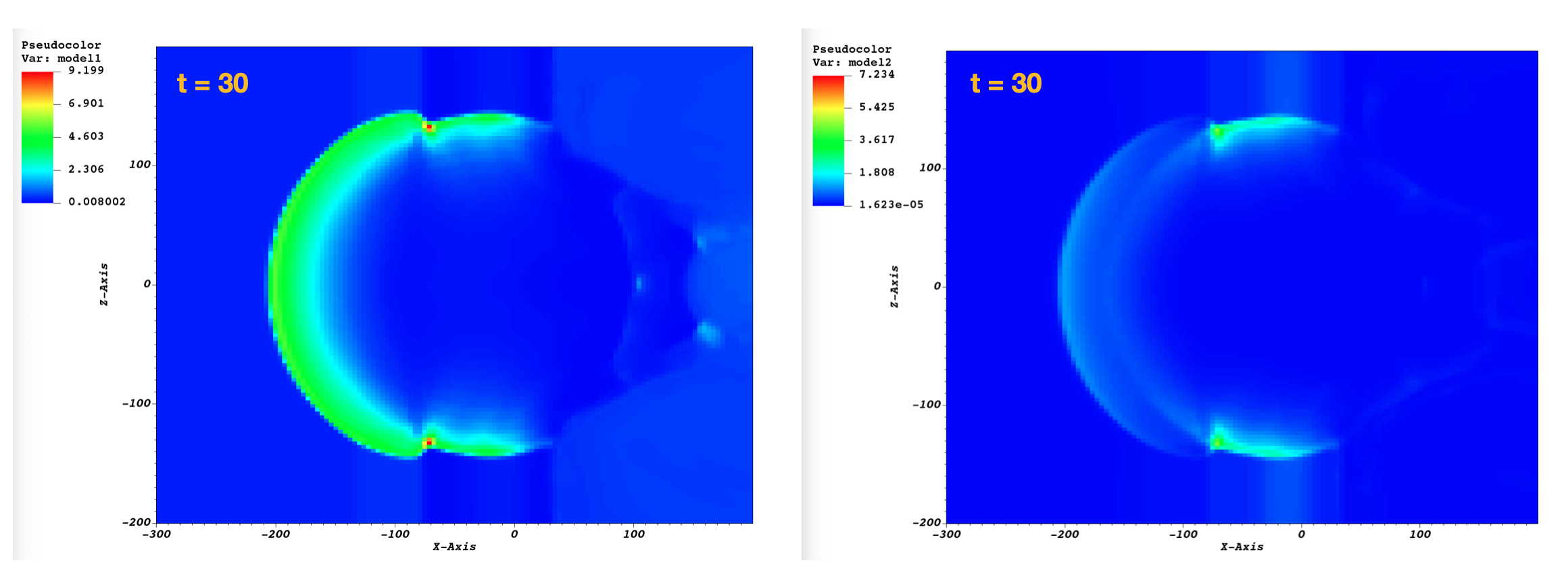}
   \caption{Slices in the xz plane of the low-resolution MHD simulation of the interaction of radio blob with a dense shell and then ISM at $t = 30$. Colors in the left panel indicate map1 of synchrotron emissivity defined as $\propto p^{1.875}$ and those in the right panel indicate map2 of synchrotron emissivity defined as $\propto pB^{1.75}$.}
   \label{fig:shell_maps}
\end{figure*}

\begin{figure}
   \includegraphics[width=\columnwidth]{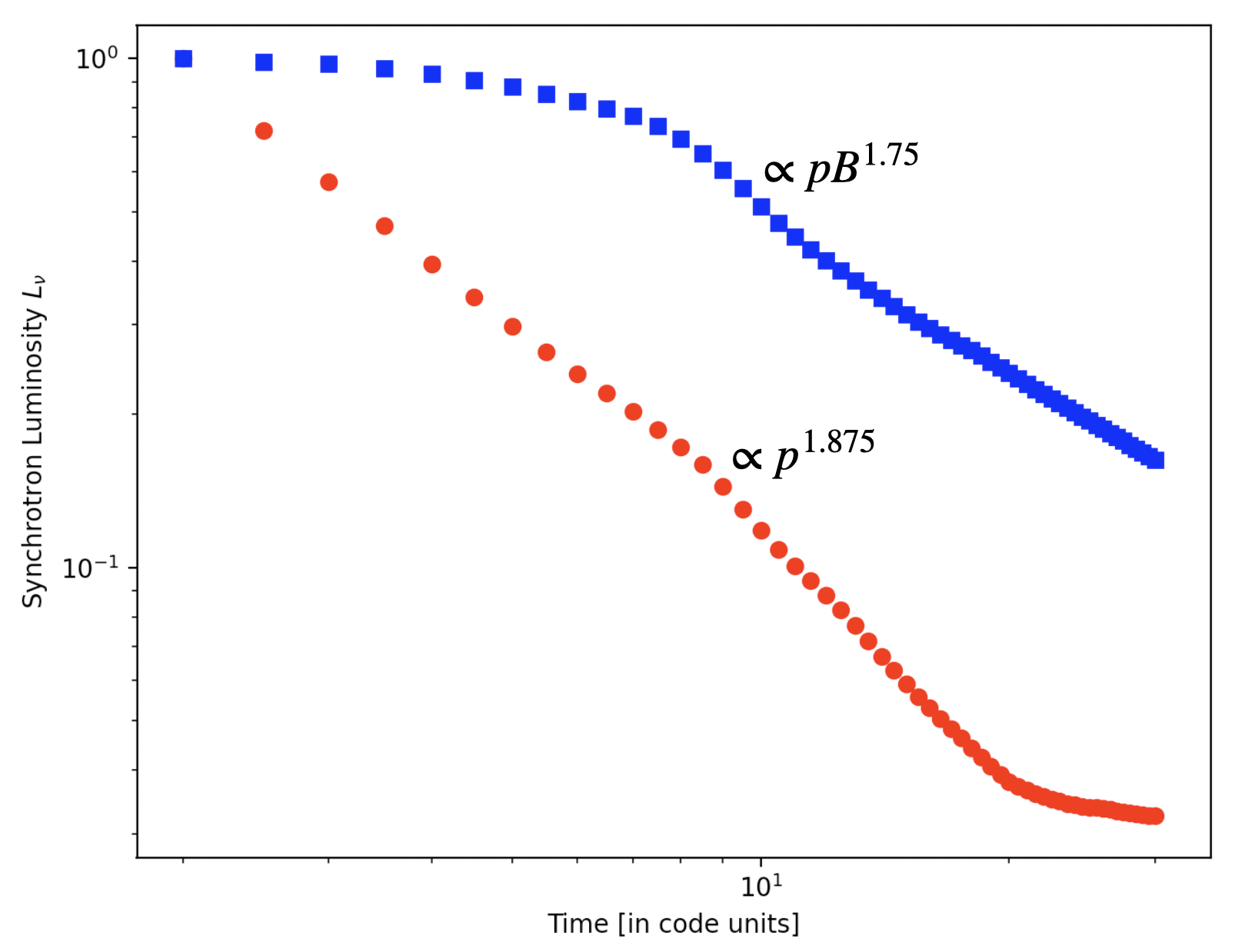}
    \caption{Normalized synthetic light curves for the two maps of figure \ref{fig:shell_maps} in log-log scale. The light curve with magnetic field amplification (red dots) undergoes two breaks and can be fit with three power laws, whereas the light curve with only magnetic field compression and no amplification (blue squares) undergoes a single break and can be fit with two power laws (see appendix~\ref{sec:SHELL} for details).}
    \label{fig:shell_TE}
\end{figure}

The initial parameters of the blob, ISM, SGR wind, and the normalizing values are the same as described in section~\ref{sec:num_set} along with shell pressure $p_{\text{shell}}=p_{\text{ISM}}$, shell density $\rho_{\text{shell}}=4\rho_{\text{ISM}}$ and \mathbfit{$B_\text{shell}$} = 4\mathbfit{$B_\text{ISM}$}. The size of the domain is $x \in [-6r_0, 4r_0]$, $y \in [-4r_0, 4r_0]$ and $z \in [-4r_0, 4r_0]$ where $r_0$ is the radius of the blob. Uniform resolution is used throughout the computational domain with total number of cells $N_{\rm X}=N_{\rm Y}=N_{\rm Z}$ = 128. The ISM extends from $-6r_0$ to $-1.4r_0$, shell size = $2r_0$ extending from $-1.4r_0$ to $0.6r_0$, and low-density cavity extends from $0.6r_0$ to $4r_0$ along the $x$ direction. We capture the dynamics of the interaction of the magnetic blob with the thick, dense shell and then the ISM from $t=0$ to $t=30$.
 % We present the synthetic synchrotron emissivity maps along with light curves in sections~\ref{sec:shell_emiss} and~\ref{sec:shell_emiss_TE}.
Results of the simulation - synthetic synchrotron emissivity maps and light curves are depicted in figures \ref{fig:shell_maps} and \ref{fig:shell_TE}.

%\subsection{Initial set-up}
%\label{sec:setup_shell}

%\subsection{Synthetic synchrotron emissivity maps}
%\label{sec:shell_emiss}

Following our procedure outlined in section~\ref{sec:emiss_maps}, we employ the two models of \citet{Reynolds2017} to create synthetic synchrotron emissivity maps in the presence of a pre-existing shell. The left and right panels of figure \ref{fig:shell_maps} are 2-D ($xz$ plane) projections of map1 and map2 of synchrotron emissivity scaled as $p^{1.875}$ and $pB^{1.75}$, respectively, at $t = 30$. The dependence of emissivity on magnetic field in the right panel (no magnetic field amplification, only compression) causes the shock to be significantly weaker than the left panel scaling only with pressure (magnetic field amplification) as it emerges out of the shell, indicating that the shell might play an important role at times much longer than the shock's shell-crossing time. 

%\subsection{Synthetic light curves}
%\label{sec:shell_emiss_TE}

We show synthetic light curves for both models in figure \ref{fig:shell_TE} from $t=5$ to $t=30$ in steps of 0.5 in log-log scale - red dots are numerical results for the box-summed normalized $p^{1.875}$ model and blue squares are numerical results for the box-summed normalized $pB^{1.75}$ model. It is clear that the late-time evolution of luminosities undergoes a break compared to early times as we show by fitting broken power laws to both light curves. 

The light curve for the case of magnetic field amplification, $L_{\nu,\text{map1}} \propto p^{1.875}$ can be fit with three power laws: at early times $\sim 2<t<8.5$, the blob's interaction with the dense shell is characterized by a steep decay in luminosity $\sim t^{-1.2}$, followed by an even steeper decay $\sim t^{-1.7}$ between $t=9$ to $t=20.5$ as the shock propagates through the shell and finally breaks out. At late times $t>20$, as effects of the shell weaken after the shock crosses the shell, luminosity undergoes a shallow decay, approaching a steady time evolution $\sim t^{-0.3}$. It is important to note that the late-time flattening of the light curve is due to the wind-ISM shock and not due to the blob-ISM shock.

The light curve for the case of magnetic field compression and no amplification, $L_{\nu,\text{map2}} \propto pB^{1.75}$ undergoes only a single break and can be fit with two power laws: at early times $\sim 2<t<7$, the blob's interaction with the magnetic field of the dense shell is characterized by a shallow decay in luminosity $\sim t^{-0.2}$, followed by a steep decay $\sim t^{-1}$ at late times $t>7$ as the shock breaks out after crossing the shell and becomes significantly weaker. It is clear that effect of magnetic field is important and causes the shell to play a dominant role in the evolution of synchrotron emission.

Our analysis of the shock dynamics in the presence of a pre-existing shell indicates that although the shell might produce a break in the radio synchrotron emission at the shock's shell crossing time, it does not account for rebrightening as reported in observations. At very late times, we expect the effects of shell to taper down significantly, thus causing the light curves to evolve similarly as the case without a shell. The shell might cause rebrightening in some regime with the appropriate shell thickness, density, and magnetic field. Hence, we are unable to conclusively eliminate this possibility. Investigation of appropriate regimes and shell parameters that might cause a flux increase can be a subject for future work.
%%%%%%%%%%%%%%%%%%%%%%%%%%%%%%%%%%%%%%%%%%%%%%%%%%

% Don't change these lines
\bsp % typesetting comment
\label{lastpage}
\end{document}